\documentclass[journal]{IEEEtran}
\usepackage{graphics}
\usepackage{colortbl}
\usepackage{multicol}
\usepackage{diagbox}
\usepackage{lipsum}
\usepackage{color}
\usepackage[cmex10]{amsmath}
\usepackage{amsthm}
\usepackage{amssymb}
\usepackage{mathrsfs}
\usepackage{mathtools}
\usepackage{amsbsy}
\usepackage[colorlinks=true,bookmarks=false,citecolor=blue,urlcolor=blue]{hyperref}
\usepackage{xcolor}
\usepackage{mathrsfs}
\usepackage{setspace}
\usepackage{float}
\usepackage{graphicx}
\usepackage{pstool}
\usepackage{epstopdf}
\usepackage{lettrine}
\usepackage{psfrag}
\usepackage{newclude}
\usepackage[normalem]{ulem}
\usepackage{latexsym}
\usepackage{algpseudocode}
\usepackage{algorithm,algpseudocode}
\usepackage{algorithmicx}
\usepackage{gensymb}
\usepackage{marginnote}
\usepackage{multirow}
\usepackage{amsmath}
\usepackage{amsmath,bm}
\usepackage{wasysym}
\usepackage{mathrsfs}
\usepackage{amsmath,cite,amsfonts,amssymb,color}
\usepackage[T1]{fontenc}
\usepackage{graphicx,psfrag}
\usepackage{pstool}
\usepackage{algorithm}
\usepackage{algpseudocode}
\usepackage{comment}
\usepackage{tabularx}
\usepackage{caption}
\usepackage{csquotes}
\usepackage{float}
\usepackage{lettrine}
\usepackage{makecell}
\usepackage{textcomp, gensymb}

\ifCLASSOPTIONcompsoc
\usepackage[caption=false,font=normalsize,labelfon
t=sf,textfont=sf]{subfig}
\else
\usepackage[caption=false,font=footnotesize]{subfig}
\fi
\allowdisplaybreaks

\graphicspath{{figures/}}
\allowdisplaybreaks

\newcommand{\RNum}[1]{\uppercase\expandafter{\romannumeral #1\relax}}

\usepackage{mathtools}

\begin{document}
\title{Resource Allocation of STAR-RIS Assisted  Full-Duplex Systems}
\author{\IEEEauthorblockN{Mohammad Reza Kavianinia\thanks{Mohammad Reza Kavianinia and Mohammad Javad Emadi are with the Department of Electrical Engineering, Amirkabir University of Technology (Tehran Polytechnic), Tehran, Iran (E-mails: \{mrezakaviani,mjemadi\}@aut.ac.ir).} and Mohammad Javad Emadi
}}   
% make the title area
\maketitle
%%====> Abstract <===%%
\begin{abstract}
In this paper, simultaneously transmitting and reflecting  reconfigurable intelligent surface (STAR-RIS) effects on the performance of full-duplex communication systems is investigated with the presence of full-duplex users, wherein the base station (BS) and the uplink users are subject to maximum transmission power constraints. Firstly, the weighted sum-rate (WSR) is derived as a system performance metric. Then, the resource allocation design is reformulated into an equivalent weighted minimum mean-square-error form and then transformed into several convex sub-problems to maximize the WSR as an optimization problem which jointly optimizes the beamforming and the combining vectors at the BS, the transmit powers of the uplink users, and phase shifts of STAR-RIS. Although the WSR optimization is non-convex, an efficient iterative alternating procedure is proposed to achieve a sub-optimal solution for the optimization problem. Secondly, the STAR-RIS's phase shifts are optimized via the successive convex approximation technique. Finally, numerical results are provided to explain how STAR-RIS improves the performance metric with the presence of full-duplex users. 
\end{abstract}
\IEEEpeerreviewmaketitle
%%====> 1. Introduction <===%%
\section{Introduction}
\lettrine{R}{ecently}, as moving from the fifth-generation (5G) networks to the beyond fifth-generation (B5G) and the sixth-generation (6G) wireless networks, huge investigations are currently being made in both academia and industry to gain more stringent requirements such as extremely high spectrum and  energy efficiency, ultra-high data rate, microsecond latency, and full dimensional coverage and connectivity \cite{saad2019vision,zhang20196g}. One of the promising new specimen of the new technologies to fulfill such requirements is reconfigurable intelligent surfaces (RIS) \cite{wu2019towards,wu2019intelligent}. By employing a large number of passive elements, an RIS intelligently controls the phase shifts of incident signals to meet the requirements of various wireless applications \cite{basar2019wireless}. RIS can dynamically alter wireless channels to improve system performance by tuning some reconfigurable elements properly \cite{mu2021capacity}.  
Nonetheless, reflecting-only RIS imposes a strong constraint on the location of nodes since the transmitter and the receiver need to be located on the same side of RIS. This constraint limits the application of RIS significantly \cite{zeng2021reconfigurable}.\\
Recently, due to the expansion of the reflective-transmissive meta-surfaces \cite{zhang2021intelligent}, a novel technique called simultaneous transmission and reflection reconfigurable intelligent surface (STAR-RIS) was proposed, where the most noteworthy advantage is that it can split the incident signal into two parts, thus can achieve ${360}^{o}$ coverage \cite{liu2021star}. Well-designed STAR-RIS, which extends the half-space coverage to full-space coverage, incurs wireless communication environments to be smart and reconfigurable. Particularly, the signal reflected in the region in front of STAR-RIS is referred to as the reflected signal, while the signal transmitted to the region behind the STAR-RIS is referred to as the transmitted signal. The two aforesaid regions are named as the reflection (R) region and transmission (T) region, respectively \cite{zhang2020beyond}.\\
Unlike reflecting-only RIS, STAR-RIS is capable of obtaining full-space coverage, as it can transmit and reflect the signal, simultaneously. By tuning the electromagnetic characteristics of the elements, the transmitting and reflecting coefficients (TRCs) can be adjusted the transmitted and reflected signals \cite{xu2021star}. In addition, three practical protocols, namely, energy splitting (ES), mode selection (MS), and time splitting (TS), have been  proposed to coordinate the T and R modes in a STAR-RIS \cite{liu2021star}. Owing to the numerous potential practical applications of
RIS deployments, RIS-assisted wireless communication systems have received increasing attention from academia to
investigate its fundamental limitations as well as enabling practical design.\\
In \cite{saeidi2021weighted}, the performance of multi-user full-duplex (FD) systems with RIS under some hardware impairments is studied, and a joint optimization problem is formulated to maximize the uplink-downlink weighted sum-rate. Besides, multiple-input multiple-output (MIMO) RIS-assisted systems are studied in \cite{niu2021weighted}, and the weighted sum-rate was maximized by optimizing the precoding matrices and TRCs alternately based on the ES scheme. In addition, the weighted sum secrecy rate by jointly designing the beamforming and the TARCs is considered in \cite{niu2021simultaneous}. RIS is also applicable to FD wireless communication systems in which transceivers are allowed to transmit and receive simultaneously in the same frequency band to increase the spectral efficiency compared to traditional half-duplex systems. This improvement is at the cost of  introducing string self-interference \cite{skouroumounis2018heterogeneous,sun2018robust}.\\
To utilize effective FD communication, some resource allocation schemes for a single IRS-assisted cognitive network have been proposed in \cite{xu2020resource}, to maximize the sum-rate of the secondary network while controlling interference leakage on the primary users. Moreover, in \cite{wang2022transmit}, the minimization of the total transmit power subject to a given minimum data rate requirement of STAR-RIS aided FD communication systems is studied. In addition, the energy efficiency maximization of STAR-RIS assisted FD communications is investigated in \cite{guan2022energy} and in \cite{perera2022sum} the sum rate performance of STAR-RIS assisted FD communication systems was studied. \\
In our work, we investigate the impact of employing STAR-RIS on the FD communication systems in the presence of FD users. It is worth mentioning that the presence of FD users alongside employing STAR-RIS operating in a FD communication system has not been considered in the previous works on STAR-IRS-assisted communications \cite{wang2022transmit,guan2022energy,perera2022sum}. 
 In this paper, to discover cost- and performance-efficient FD user systems, an FD STAR-IRS-assisted system is investigated to enhance system performance, while assuming FD users to figure out how the FD STAR-IRS-assisted system
behaves in the presence of FD users. It is considered that STAR-RIS exists in the network to cooperatively support the uplink (UL) and the downlink (DL) modes while interacting with a multi-antenna BS. Since the UL and DL communications are operated in an FD manner, not only does the signal of the UL users cause interference to the other DL user, but also the BS is also subject to a non-negligible self-interference.
Therefore, our purpose is to
design efficient resource allocation algorithms for maximizing the weighted sum-rate (WSR). In the following, the contributions of this paper are summarized.
\begin{itemize}
    \item The achievable rates of the uplink and the downlink FD users are derived, while the maximization problems for the UL-DL weighted system sum-rate are formulated. To maximize the aforementioned performance metric, we jointly optimize the beamforming vector for the downlink users subject to the maximum power constraint at the BS and the combining vector of the uplink users at the BS. Moreover, the optimization problem for  three practical protocols (ES, MS, TS) of STAR-RIS is formulated. 
    \item Since the aforementioned optimization problem is not jointly concave with respect to the optimization parameters, a suboptimal algorithm based on the iterative alternating optimization approach is proposed. particularly, for the given STAR-RIS’s phase shift matrices, we reformulate the optimization problem into an equivalent weighted minimum mean square error (WMMSE) problem to achieve the DL beamformer, the UL combining vector, and the UL users’ transmit powers, iteratively. Thus, for the given beamformer, the combiner and the power allocation solutions, we solve the STAR-RIS’s amplitudes and phase shifts optimization problem via the successive convex approximation (SCA) technique to obtain a suboptimal solution.
    \item Our numerical results represent that employing STAR-RIS  can significantly improve the WSR performance compared with that of the conventional RIS system or without RIS or even UL-HD, DL-HD and HD schemes. Finally, it is shown that the location of STAR-RIS that causes the WSR maximization in that specific location.
\end{itemize}

\textit{Organization}: The rest of this paper is organized as follows. In Section II, the proposed system model is introduced and the WSR is derived. Section III formulates the optimization problem and provides the analyses of the proposed scheme. Numerical results are discussed in Section IV, and finally, Section V concludes the paper. \\
\indent\textit{Notations}: $\mathbb{C}^{M \times N}$ denotes the space of $M \times N$ complex valued matrices. $\boldsymbol{x}^T$ and $\boldsymbol{x}^H$ denote the transpose and conjugate transpose of a vector $\boldsymbol{x}$. Moreover, $\textrm{diag}(\boldsymbol{x})$ describes a diagonal matrix with elements of the vector $\boldsymbol{x}$ as diagonal elements. For complex-valued vector $\boldsymbol{x}$,  $\left|\boldsymbol{x}\right|$ denotes its Euclidean norm and, $\mathfrak{R}(x)$, is the
real part of $x$. The matrix $\boldsymbol{\mathrm{I}}_N$ represents a $N \times N$ identity matrix. $s \sim \mathcal{C}\mathcal{N} (0,\sigma^2)$ shows that $s$ has complex Gaussian distribution with zero mean and variance $\sigma^2$. 
%%====> 2. System Model <===%%

\section{System Model}
\begin{figure}[t!]
    \centering
    \pstool[scale=0.5]{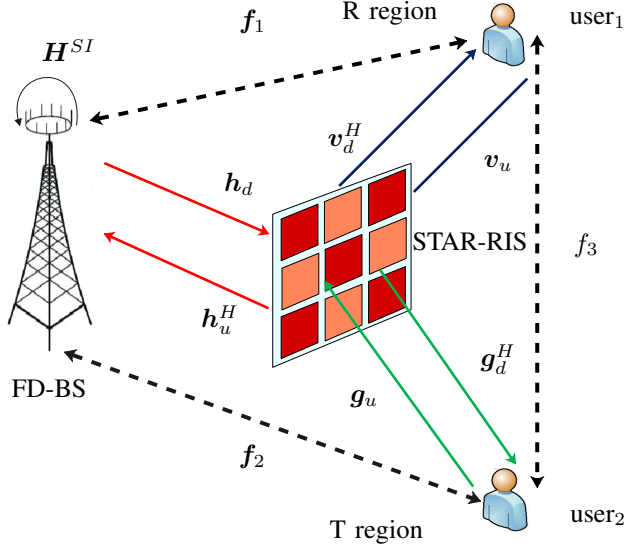}{
    \psfrag{f}{\hspace{-1.5cm}$\boldsymbol{f}_1$}
    \psfrag{B}{$\boldsymbol{f}_2$}
    \psfrag{C}{${f}_3$}
    \psfrag{D}{$\textrm{user}_1$}
    \psfrag{E}{$\textrm{user}_2$}
    \psfrag{F}{\hspace{-0.85cm} \text{STAR-RIS}}
    \psfrag{G}{$\boldsymbol{h}_d$}
    \psfrag{H}{$\boldsymbol{h}_u^{H}$}
    \psfrag{u}{$\boldsymbol{v}_d^{H}$}
    \psfrag{J}{$\boldsymbol{v}_u$}
    \psfrag{K}{$\boldsymbol{g}_d^{H}$}
    \psfrag{L}{$\boldsymbol{g}_u$}
    \psfrag{M}{\hspace{-0.50cm} \textrm{FD-BS}}
    \psfrag{N}{\hspace{-0.90cm} \textrm{R region}}
    \psfrag{O}{\hspace{-0.90cm}\textrm{T region}}
    \psfrag{P}{$\boldsymbol{H}^{SI}$}
    }
    \caption{The STAR-RIS-aided Full-Duplex system model.}
    \label{fig:1}
\end{figure}

As depicted in Fig. \ref{fig:1}, a STAR-RIS assisted FD communication system is considered in which $N_t$-antenna FD BS is intended to communicate with two single-antenna FD users with the assistance of the STAR-RIS with $M$ antenna elements. The amplitude and phase of reflected and transmitted signals at each element can be adjusted independently. The  $\textrm{user}_1$ $(U_1)$ and  $\textrm{user}_2$ $(U_2)$ are placed at reflection region and transmission region of the STAR-RIS, respectively. Since both users operate at the same frequency band, received signal at each user is interfered by the UL signal of the other user. This interference is both directly and via STAR-RIS.\\
In this paper, a flat-fading narrowband channel is considered. It is assumed that the channel state information of both users is available at the BS which results to an upper-bound in the performance of a realistic system as discussed in  \cite{wu2019intelligent,xu2020resource,pan2020multicell}. Note that in the  considered system, CSI can be estimated using the reciprocity principle since the same frequency band is used for UL and DL.
%%====> A. Signal Model of STAR-RIS <===%%
\subsection{Signal Model of STAR-RIS } 
Signal coming from a given element of STAR-RIS is divided into  a transmitted and a reflected signal. Assume that $s_m$ denotes the incident signal on the $m$-th element of STAR-RIS, where $m \in \mathcal{M}  \triangleq \{ 1, 2, ..., M\}$. Then, the transmitted and reflected signals can be formulated as $t_m =\sqrt{\beta_m^t}e^{j\phi_m^t}s_m$  and $r_m = \sqrt{\beta_m^r}e^{j\phi_m^r}s_m$, respectively, where $\sqrt{\beta_m^t} \in \left [0,1\right] $, $\phi_m^t \in \left [0,2\pi\right] $ and $\sqrt{\beta_m^r} \in \left [0,1\right]$, $\phi_m^r \in \left [0,2\pi\right]$ are the amplitude and phase shift response of the $m$-th element, respectively. For each element, the phase shifts for transmission and reflection (i.e., $\phi_m^t$ and $\phi_m^r$ ) can be independently set. However,  the amplitude components of transmission and reflection (i.e., $\sqrt{\beta_m^t}$ and $\sqrt{\beta_m^r}$) are coupled by the law of energy conservation and the sum of the energies of the transmitted and reflected signals has to be equal to the energy of the incident signal, i.e., $|t_m|^2 + |r_m|^2 = |s_m|^2 $, which  implies $\beta_m^t + \beta_m^r = 1, \forall m \in \mathcal{M}$. 
%%====> B. STAR-RIS's Three Practical Protocols <===%%
\subsection{STAR-RIS's Three Practical Protocols}
In this part, three known protocols of STAR-RIS, namely ES, MS, and TS  are briefly explained.
\subsubsection{Energy Splitting}In the ES protocol, all the elements of the STAR-RIS work in transmitting and reflecting modes simultaneously. Hence, the TARCs are modelled as $\boldsymbol{\Phi}_t^{ES}=\textrm{diag}(\sqrt{\beta_1^t}e^{j\phi_1^t}, \dots , \sqrt{\beta_M^t}e^{j\phi_M^t})$, and $\boldsymbol{\Phi}_r^{ES}=\textrm{diag}(\sqrt{\beta_1^r}e^{j\phi_1^r}, \dots , \sqrt{\beta_M^r}e^{j\phi_M^r})$, where $\beta_m^t,\beta_m^r \in \left[0,1\right]$, $\beta_m^t + \beta_m^r =1$, and $\phi_m^t,\phi_m^r \in \left[0,2\pi\right), \forall m\in\mathcal{M}$. ES provides a high system flexibility since the transmitting and reflecting coefficients of every single element can be optimized. Note that this flexibility is at the cost of larger overhead for exchanging information between BS and the STAR-RIS.  
\subsubsection{Mode Selection}In the MS protocol, all elements of STAR-RIS are dedicated to two parts, so that one part with $M_t$  elements operating in transmitting mode, while another part with $M_r$ elements operating in the reflecting mode, as such satisfying $M_t + M_r=M$. The TRACs of this protocol are given by $\boldsymbol{\Phi}_t^{MS}=\textrm{diag}(\sqrt{\beta_1^t}e^{j\phi_1^t}, \dots , \sqrt{\beta_M^t}e^{j\phi_M^t})$, and $\boldsymbol{\Phi}_r^{MS}=\textrm{diag}(\sqrt{\beta_1^r}e^{j\phi_1^r}, \dots , \sqrt{\beta_M^r}e^{j\phi_M^r})$, respectively, where $\beta_m^t,\beta_m^r \in \{0,1\}, \beta_m^t + \beta_m^r =1$, and $\phi_m^t,\phi_m^r \in \left[0,2\pi\right), \forall m\in\mathcal{M}$. Note that the on-off concept in MS is different that the on-off operation in tunable elements (e.g. pin diods) in the RIS.
\subsubsection{Time Switching} The elements switch between the two modes in different time slots. The TARCs are given by $\boldsymbol{\Phi}_t^{TS}=\textrm{diag}(e^{j\phi_1^t}, \dots , e^{j{\phi_M^t}})$ and $\boldsymbol{\Phi}_r^{TS}=\textrm{diag}(e^{j\phi_1^r}, \dots , e^{j\phi_M^r})$, respectively, where, $\phi_m^t,\phi_m^r \in \left[0,2\pi\right), \forall m \in \mathcal{M}$. For TS protocol, T and R coefficients can be designed independently. However, time synchronization is needed for switching between two modes, which leads to a higher complexity design.       
\subsection{Signal Transmissions and Receptions}
The FD-BS transmits the precoded signal $\boldsymbol{x}^{DL}= \sum_{k\in \mathcal{K}} \boldsymbol{\mathit{w}}_k s_k $ to the ${\mathcal{K} \in \{1,2\}}$, where ${\mathcal{K}}$ denotes signal of $\mathit{k}$-th user, $s_k \sim \mathcal{C}\mathcal{N} (0,1)$ and $\boldsymbol{\mathit{w}}_k \in \mathbb{C}^{N_t \times 1 } $ is the BS transmit beamforming vector. The $\mathit{l}$-th UL user transmits $x_l^{UL}=\sqrt{\rho_l}q_l$, in which $ \mathcal{L} \in \{1,2\}$,  $q_l \sim \mathcal{C}\mathcal{N} (0,1)$ is the information symbol and $\rho_l$ is the UL power of the $\mathit{l}$-th user. The BS  transmits signals to $U_1$ and $U_2$ and receives signals  from $U_1$ and $U_2$, synchronously. The $U_1$ receives 
\begin{equation}
    \begin{split}
            y_1^{DL}&=\Big(\boldsymbol{v}_d^{H}\boldsymbol{\Phi}_r\boldsymbol{h}_d+\boldsymbol{f}_1\Big)\sum_{k\in \mathcal{K}}\boldsymbol{\mathit{w}}_k s_k +\Big(\boldsymbol{v}_d^{H}\boldsymbol{\Phi}_t\boldsymbol{g}_u+{f}_3\Big)\sqrt{\rho_2}q_2\\
            &+ n_{U_1}\label{eq:1},
            \raisetag{13pt}
    \end{split}
\end{equation}
and the $U_2$ receives the following signal\vfill  
\begin{equation}
    \begin{split}
        y_2^{DL}&=\Big(\boldsymbol{g}_d^{H}\boldsymbol{\Phi}_t\boldsymbol{h}_d+\boldsymbol{f}_2\Big)\sum_{k\in \mathcal{K}}\boldsymbol{\mathit{w}}_ks_k +\Big(\boldsymbol{g}_d^{H}\boldsymbol{\Phi}_t\boldsymbol{v}_u+{f}_3\Big)\sqrt{\rho_1}q_1\\
        &+ n_{U_2}\label{eq:2},
        \raisetag{13pt}
    \end{split}
\end{equation}
where $\boldsymbol{v}_d^{H} \in \mathbb{C}^{1 \times M}$ is the channels between  the STAR-RIS and $U_1$, $\boldsymbol{h}_d \in \mathbb{C}^{M \times N_t}$ is the channels between the BS and the STAR-RIS, $\boldsymbol{f}_1 \in \mathbb{C}^{1 \times N_t}$ is the channels between the BS and $U_1$, $\boldsymbol{g}_u \in \mathbb{C}^{M \times 1}$ is the channels between $U_2$ and the STAR-RIS, ${f}_3 \in \mathbb{C}$ is the channel between $U_1$ and $U_2$, $\boldsymbol{g}_d^{H} \in \mathbb{C}^{1 \times M}$ is the channels between the STAR-RIS and $U_2$, $\boldsymbol{f}_2 \in \mathbb{C}^{1 \times N_t}$ is the channels between the BS and $U_2$ and $\boldsymbol{v}_u \in \mathbb{C}^{M \times 1}$ is the channels between $U_1$ and the STAR-RIS. Besides, diagonal matrices $\boldsymbol{\Phi}_r,\boldsymbol{\Phi}_t \in \mathbb{C}^{M \times M}$ express amplitude and phase shift of STAR-RIS. furthermore, $n_{U_1} \sim \mathcal{C}\mathcal{N} (0,\sigma_{U_1}^2)$ and $n_{U_2} \sim \mathcal{C}\mathcal{N} (0,\sigma_{U_2}^2)$ model the additive white Gaussian noise (AWGN) at the DL users. The received signal at the BS is given by  
\begin{equation}
\begin{split}
        \boldsymbol{y}^{UL}&=\Big(\boldsymbol{h}_u^{H}\boldsymbol{\Phi}_r
                             \boldsymbol{v}_u +\boldsymbol{f}_1^{H}\Big)\sqrt{\rho_1}q_1 +\Big(\boldsymbol{h}_u^{H}\boldsymbol{\Phi}_t\boldsymbol{g}_u +   \boldsymbol{f}_2^{H}\Big)\sqrt{\rho_2}q_2\\
                             &+\boldsymbol{H}^{SI}\boldsymbol{x}^{DL} + \boldsymbol{n}^{UL}\label{eq:3},
                             \raisetag{13pt}
\end{split}
\end{equation}
where $\boldsymbol{h}_u^{H} \in \mathbb{C}^{N_t \times M} $, $\boldsymbol{v}_u \in \mathbb{C}^{M \times 1}$ and $\boldsymbol{g}_u \in \mathbb{C}^{M \times 1}$ denote the channels between the STAR-RIS and the BS, the channels between $U_1$ and the STAR-RIS and the channels between $U_2$ and STAR-RIS, respectively, and the term $\boldsymbol{H}^{SI}\boldsymbol{x}^{DL}$ indicates the residual self-interference (RSI) \cite{nguyen2016optimal}. Similar to \cite{ngo2014multipair}, we assume that $\boldsymbol{H}^{SI}$ is unknown at the BS and each element has i.i.d. complex zero-mean Gaussian distribution with variance $\hat{\sigma}^2$, and $\boldsymbol{n}^{UL} \sim \mathcal{C}\mathcal{N} (0,\sigma_{UL}^2\mathbf{I}_{N_t}) $ models AWGN at the BS.
\subsection{Weighted Sum-Rate Formulation}
 Here, some achievable rates of the users in DL and UL modes are derived and WSR is presented. It can be shown the following achievable rate in bits per channel use (bpcu) can be achieved by $U_1$ and $U_2$ in DL 
 \begin{equation}
    R_1^{DL}= \log_2(1+\gamma_1^{DL}) \hspace{0.25cm} \textrm{[bpcu]},
 \end{equation} 
 \begin{equation}
    R_2^{DL}= \log_2(1+\gamma_2^{DL}) \hspace{0.25cm} \textrm{[bpcu]},
 \end{equation}
  where $\gamma_1^{DL}$ and $\gamma_2^{DL}$ are DL signal-to-interfrence-plus-noise-ratio (SINR), which are calculated as follows. First for the sake of brevity, let's define $\boldsymbol{H}_{r1}=\boldsymbol{v}_d^{H}\boldsymbol{\Phi}_r\boldsymbol{h}_d+\boldsymbol{f}_1$   
  , $\boldsymbol{G}_{t1}=\boldsymbol{v}_d^{H}\boldsymbol{\Phi}_t\boldsymbol{g}_u+{f}_3$
  , $\boldsymbol{H}_{r2}=\boldsymbol{g}_d^{H}\boldsymbol{\Phi}_t\boldsymbol{h}_d+\boldsymbol{f}_2$
  , $\boldsymbol{G}_{t2}=\boldsymbol{g}_d^{H}\boldsymbol{\Phi}_t\boldsymbol{v}_u+{f}_3$
  , $\boldsymbol{H}_{r3}=\boldsymbol{h}_u^{H}\boldsymbol{\Phi}_r\boldsymbol{v}_u+\boldsymbol{f}_1^{H}$  and $\boldsymbol{G}_{t3}=\boldsymbol{h}_u^{H}\boldsymbol{\Phi}_t\boldsymbol{g}_u + \boldsymbol{f}_2^{H}$. Then, $\gamma_1^{DL}$ and $\gamma_2^{DL}$ are given by 
 \begin{equation}
    \gamma_1^{DL}=\frac{{\big|\boldsymbol{H}_{r1}\boldsymbol{\mathit{w}}_1 \big|}^2}{{\big|\boldsymbol{G}_{t1}\sqrt{\rho_2} \big|}^2+{\big|\boldsymbol{H}_{r1}\boldsymbol{\mathit{w}}_2 \big|}^2 + \sigma_{U_1}^2}\label{eq:6},
 \end{equation}
 and
 \begin{equation}
    \gamma_2^{DL}=\frac{{\big|\boldsymbol{H}_{r2}\boldsymbol{\mathit{w}}_2 \big|}^2}{{\big|\boldsymbol{G}_{t2}\sqrt{\rho_1} \big|}^2+{\big|\boldsymbol{H}_{r2}\boldsymbol{\mathit{w}}_1 \big|}^2 + \sigma_{U_2}^2}\label{eq:7}.
 \end{equation}\vfill
The BS applies the combining vector $\boldsymbol{U}_l \in \mathbb{C}^{N_t \times 1}$ to recover the data symbol of $l$-th UL user, as $\hat{q}_l=\boldsymbol{U}_l^{H}\boldsymbol{y}^{UL}$. Therefore, the UL achievable rates for $U_1$ and $U_2$ can be calculated as
 \begin{equation}
    R_1^{UL}= \log_2(1+\gamma_1^{UL}) \hspace{0.25cm} \textrm{[bpcu]},
 \end{equation}
 \begin{equation}
    R_2^{UL}= \log_2(1+\gamma_2^{UL}) \hspace{0.25cm} \textrm{[bpcu]},
 \end{equation}
where $\gamma_1^{UL}$ and $\gamma_2^{UL}$ are UL SINR and are given in \eqref{eq:11} and \eqref{eq:12} in the following, respectively. We assume average RSI power similar to \cite{mohammadi2018uplink} to simplify the effect of RSI. Therefore, the average RSI power at BS for the $l$-th user is given  by
\begin{equation}
            \textrm{RSI}(\boldsymbol{U}_l)=\mathbb{E}\Big\{ {\big| \boldsymbol{U}_l^{H}\boldsymbol{H}^{SI}\boldsymbol{x}^{DL} \big|}^2 \Big\}
            =\hat{\sigma}^2{\big| \boldsymbol{U}_l \big|}^2\sum_{k\in \mathcal{K}}{\big| \boldsymbol{\mathit{w}}_k \big|}^2 
\end{equation}
thus, $\gamma_1^{UL}$ and $\gamma_2^{UL}$ are given by
\begin{equation}
    \gamma_1^{UL}=\frac{{\big|\boldsymbol{U}_1^{H}\boldsymbol{H}_{r3}\sqrt{\rho_1} \big|}^2}{{\big|\boldsymbol{U}_1^{H}\boldsymbol{G}_{t3}\sqrt{\rho_2} \big|}^2+\boldsymbol{\textrm{RSI}}(\boldsymbol{U}_1)+ {\big|\boldsymbol{U}_1^{H}\big|}^2\sigma_{UL}^2}\label{eq:11},
 \end{equation}
 and 
 \begin{equation}
    \gamma_2^{UL}=\frac{{\big|\boldsymbol{U}_2^{H}\boldsymbol{G}_{t3}\sqrt{\rho_2} \big|}^2}{{\big|\boldsymbol{U}_2^{H}\boldsymbol{H}_{r3}\sqrt{\rho_1} \big|}^2+\boldsymbol{\textrm{RSI}}(\boldsymbol{U}_2)+ {\big|\boldsymbol{U}_2^{H}\big|}^2\sigma_{UL}^2}\label{eq:12}.
 \end{equation}
On the other hand, for the TS scheme, since there is no inner and user interference, the achievable rates are
 \begin{equation}
    R_1^{DL}= \tau_1^{DL}\log_2\Bigg(1+\frac{{\big|\boldsymbol{H}_{r1}\boldsymbol{\mathit{w}}_1 \big|}^2}{\sigma_{DL}^2}\Bigg)  \hspace{0.9cm},
 \end{equation}
 \begin{equation}
    R_2^{DL}=\tau_2^{DL}\log_2\Bigg(1+\frac{{\big|\boldsymbol{H}_{r2}\boldsymbol{\mathit{w}}_2 \big|}^2}{\sigma_{DL}^2}\Bigg)  \hspace{0.9cm},
 \end{equation}
 \begin{equation}
    R_1^{UL}= \tau_1^{UL}\log_2\Bigg(1+\frac{{\big|\boldsymbol{U}_1^{H}\boldsymbol{H}_{r3}\sqrt{\rho_1} \big|}^2}{ {\big|\boldsymbol{U}_1^{H}\big|}^2\sigma_{UL}^2}\Bigg)  \hspace{0.25cm} ,
 \end{equation}
 \begin{equation}
    R_2^{UL}= \tau_2^{UL}\log_2\Bigg(1+\frac{{\big|\boldsymbol{U}_2^{H}\boldsymbol{G}_{t3}\sqrt{\rho_2} \big|}^2}{ {\big|\boldsymbol{U}_2^{H}\big|}^2\sigma_{UL}^2}\Bigg)  \hspace{0.25cm},
 \end{equation}
 and, the WSR is calculated as 
 \begin{equation}
    \textrm{WSR}=\alpha_1\sum_{k\in \mathcal{K}}R_k^{DL}+\alpha_2\sum_{l\in \mathcal{L}}R_l^{UL},
 \end{equation}
 where $\alpha_1 \ge 0$ and  $\alpha_2 \ge 0$ control the weights of sum-rate at DL and the UL, respectively.  
%%====> 3. Optimization Problem Formulation <===%%
\section{Optimization Problem Formulation} \label{Sec:Sec3}
We aim to maximize the WSR by jointly optimizing the BS
transmit beamforming, the transmission power of users, the combining vector and TARCs. For the ES scheme, the problem is formulated as   
\begin{subequations}
\label{P1}
\begin{align}
\textrm{ES:}\max_{\substack{\boldsymbol{w}_k,\rho_l,\boldsymbol{U}_l,\\\boldsymbol{\Phi}_t, \boldsymbol{\Phi}_r}}
\quad & \alpha_1\sum_{k\in \mathcal{K}}R_k^{DL}+\alpha_2\sum_{l\in \mathcal{L}}R_l^{UL}\label{eq:18a}\\
\textrm{s.t.} \quad & \sum_{k\in \mathcal{K}} {\big| \boldsymbol{w}_k \big|}^2 \le P_{max}^{BS},\label{eq:18b}\\
 \quad & \rho_l \le P_{max}^{l}, \forall  l, \label{eq:18c}  \\
 \quad & \big[\boldsymbol{\Phi}_x \big]_m = \sqrt{\beta_m^x}e^{j\phi_m^x}, \forall  m \in \mathcal{M},  \label{eq:18d} \\ 
 \quad & \beta_m^x \in \big[0,1 \big], \phi_m^x \in \big[0,2\pi \big), \forall  x \in \big\{t,r\big\},\label{eq:18e} \\
 \quad & \beta_m^t + \beta_m^r = 1,\forall  m \in \mathcal{M}, \label{eq:18f}
\end{align}
\end{subequations}
where \eqref{eq:18b} denotes the maximum power constraint with the maximum transmit power $P_{max}^{BS}$ at the BS, \eqref{eq:18c} indicates the maximum power constraint of each UL user with $P_{max}^{l}$ as the maximum transmit power at $l$-th UL user, the constraints \eqref{eq:18d}-\eqref{eq:18f} model the phase and amplitude limitations of STAR-RIS elements.\\
It is known that \eqref{P1} can be reformulated to the MS scheme by replacing $\beta_m^x \in \big[0,1 \big]$ with $\beta_m^x \in \big\{0,1 \big\}$. Then for the MS scheme, WSR optimization becomes
\begin{subequations}
\label{P2}
\begin{align}
\textrm{MS:}\max_{\substack{\boldsymbol{w}_k,\rho_l,\boldsymbol{U}_l,\\\boldsymbol{\Phi}_t, \boldsymbol{\Phi}_r}}
\quad & \alpha_1\sum_{k\in \mathcal{K}}R_k^{DL}+\alpha_2\sum_{l\in \mathcal{L}}R_l^{UL}\\
\textrm{s.t.} \quad & \sum_{k\in \mathcal{K}} {\big| \boldsymbol{w}_k \big|}^2 \le P_{max}^{BS},\\
 \quad & \rho_l \le P_{max}^{l}, \forall  l,   \\
 \quad & \big[\boldsymbol{\Phi}_x \big]_m = \sqrt{\beta_m^x}e^{j\phi_m^x}, \forall  m \in \mathcal{M}, \\ 
 \quad & \beta_m^x \in \big\{0,1 \big\}, \phi_m^x \in \big[0,2\pi \big), \forall  x \in \big\{t,r\big\}, \label{eq:19e} \\
 \quad & \beta_m^t + \beta_m^r = 1,\forall  m \in \mathcal{M},\label{eq:19f} 
\end{align}
\end{subequations}
And finally, for the TS scheme, the WSR optimization is formulated as
\begin{subequations}
\label{P3}
\begin{align}
\begin{split}
\textrm{TS:}\max_{\substack{\boldsymbol{w}_k,\rho_l,\boldsymbol{U}_l,\boldsymbol{\Phi}_t, \boldsymbol{\Phi}_r,\\ \tau_k^{DL}, \tau_l^{UL}} }
\quad & \alpha_1\sum_{k\in \mathcal{K}}R_k^{DL}
+\alpha_2\sum_{l\in \mathcal{L}}R_l^{UL}
\end{split}\\
\textrm{s.t.} \quad &  \sum_{k\in \mathcal{K}} \tau_k^{DL}{\big| \boldsymbol{w}_k \big|}^2 \le P_{max}^{BS},\\
 \quad & \tau_l^{UL}\rho_l \le P_{max}^{l}, \forall  l,   \\
 \quad & \big[\boldsymbol{\Phi}_x \big]_m = e^{j\phi_m^x}, \forall  m \in \mathcal{M}, \\ 
 \quad &  \phi_m^x \in \big[0,2\pi \big), \forall  x \in \big\{t,r\big\}, \\
 \quad & \tau_k^{DL} \in \big[0,1 \big], \tau_l^{UL} \in \big[0,1 \big], \forall k,l ,\\
 \quad & \sum_{k\in \mathcal{K}}\tau_k^{DL} + \sum_{l\in \mathcal{L}}\tau_l^{UL}=1. 
\end{align}
\end{subequations}
The optimization problems \eqref{P1}, \eqref{P2} and \eqref{P3} are non-convex and obtaining their globally optimal solutions is not possible. Therefore, we pursue an alternative optimization method which attempts to achieve a sub-optimal solution. To do so, for given TARCs, the optimization problems are transformed into an equivalent WMMSE formulation which utilizes an iterative method with low computational complexity used in \cite{shi2011iteratively}. In the following parts, we decompose these equivalent optimization problems into a set of convex subproblems, and the beamformer, the combining vector at the BS and the transmitted power of the UL users are optimized. Eventually, for the given solutions, we optimize the TARCs matrices via the SCA technique; this procedure continues until the convergence.
\subsection{WMMSE Optimization Problem for Given $\boldsymbol{\Phi}_t$ and $\boldsymbol{\Phi}_r$ }
For a given  $\boldsymbol{\Phi}_t$ and $\boldsymbol{\Phi}_r$, by applying  a similar WMMSE framework as in \cite{saeidi2021weighted, masoumi2019performance, van2018large},  the optimization problems of \eqref{P1}, \eqref{P2} and \eqref{P3} are transformed into the following WMMSE problem 
\begin{subequations}
\label{P4}
    \begin{align}
        \begin{split}
                \mathcal{P}_1:\min_{\substack{\boldsymbol{w}_k,\rho_l,\boldsymbol{U}_l, u_k,\\ \mu_k^{DL}, \mu_l^{UL} }}
                \quad & \alpha_1\sum_{k\in \mathcal{K}}(\mu_k^{DL}e_k^{DL}-\ln \mu_k^{DL})\\ 
                & +\alpha_2\sum_{l\in \mathcal{L}}(\mu_l^{UL}e_l^{UL}-\ln \mu_l^{UL})\label{eq:21a}
        \end{split}\\
        \textrm{s.t.} \quad & \sum_{k\in \mathcal{K}} {\big| \boldsymbol{w}_k \big|}^2 \le P_{max}^{BS},\\
        \quad & \rho_l \le P_{max}^{l}, \forall  l,
    \end{align}
\end{subequations}
wherein $\mu_k^{DL}$ and $\mu_l^{UL}$ are weight factors for DL and UL, respectively. In addition, $e_k^{DL}$ is defined as 
\begin{equation}
    \begin{split}
       e_k^{DL}&=\mathbb{E}\Big\{ {\big| \hat{s}_k - s_k \big|}^2 \Big\}=\mathbb{E}\Big\{ {\big| u_ky_k^{DL} - s_k \big|}^2 \Big\}\\
       &= {\big| u_k \big|}^2\Bigg({\big| \boldsymbol{H}_{rk}\boldsymbol{\mathit{w}}_k  \big|}^2+{\big| \boldsymbol{G}_{tk}\sqrt{\rho_{\acute{k}}} \big|}^2+{\big| \boldsymbol{H}_{rk}\boldsymbol{\mathit{w}}_{\acute{k}} \big|}^2 + \sigma_{DL}^2\Bigg)\\
       &-2\mathfrak{R}\bigg(u_k\boldsymbol{H}_{rk}\boldsymbol{\mathit{w}}_k\bigg) + 1,\label{eq:22}
    \end{split}
    \raisetag{15pt}
 \end{equation}
 where $s_k$ is detected by detecting factor $u_k \in \mathbb{C}$, i.e., $\hat{s}_k=u_ky_k^{DL}$, and  $\acute{k} = 2$, if $k = 1$; and $\acute{k} = 1$, otherwise. The error cost functions of UL are given in the following,
 \begin{equation}
    \begin{split}
       e_1^{UL}&=\mathbb{E}\Big\{ {\big| \hat{q}_1 - q_1 \big|}^2 \Big\}=\mathbb{E}\Big\{ {\big| \boldsymbol{U}_1^H\boldsymbol{y}^{UL} - q_1 \big|}^2 \Big\}\\
       &= {\big| \boldsymbol{U}_1^H \big|}^2\Bigg({\big| \boldsymbol{H}_{r3}\sqrt{\rho_1}  \big|}^2+{\big| \boldsymbol{G}_{t3}\sqrt{\rho_2} \big|}^2+\hat{\sigma}^2{\big|(\boldsymbol{\mathit{w}}_1+\boldsymbol{\mathit{w}}_2)  \big|}^2\\
       &+ \sigma_{UL}^2\Bigg)
       -2\mathfrak{R}\bigg(\boldsymbol{U}_1^H\boldsymbol{H}_{r3}\sqrt{\rho_1}\bigg) + 1,\label{eq:23}
    \end{split}
    \raisetag{17pt}
 \end{equation}
 and
 \begin{equation}
    \begin{split}
       e_2^{UL}&=\mathbb{E}\Big\{ {\big| \hat{q}_2 - q_2 \big|}^2 \Big\}=\mathbb{E}\Big\{ {\big| \boldsymbol{U}_2^H\boldsymbol{y}^{UL} - q_2 \big|}^2 \Big\}\\
       &= {\big| \boldsymbol{U}_2^H \big|}^2\Bigg({\big| \boldsymbol{H}_{r3}\sqrt{\rho_1}  \big|}^2+{\big| \boldsymbol{G}_{t3}\sqrt{\rho_2} \big|}^2+\hat{\sigma}^2{\big|(\boldsymbol{\mathit{w}}_1+\boldsymbol{\mathit{w}}_2)  \big|}^2\\
       &+ \sigma_{UL}^2\Bigg)
       -2\mathfrak{R}\bigg(\boldsymbol{U}_2^H\boldsymbol{G}_{t3}\sqrt{\rho_2}\bigg) + 1.\label{eq:24}
    \end{split}
     \raisetag{17pt}
 \end{equation}
 It is known that the optimization problem $\mathcal{P}_1$ is not jointly convex but is convex in each optimization variable, which motivates the utilization of an iterative algorithm based on the alternating optimization.\\
 In order to derive the optimal values of $\boldsymbol{\mathit{w}}_k$, $\rho_l$ and $\big\{\boldsymbol{U}_l,  u_k, \mu_k^{DL}, \mu_l^{UL}\big\}$, we transform the problem $\mathcal{P}_1$ into several sub-problems as follows:
\subsubsection{Optimizing the BS Beamforming Vector}
For a given set of \big\{ $\boldsymbol{U}_l, u_k, \mu_k^{DL}, \mu_l^{UL},  \rho_l$  \big\}, optimization problem \eqref{P4} is rewritten as 
\begin{subequations}
    \begin{align}
        \begin{split}
                \mathcal{P}_{1.1}:\min_{\boldsymbol{w}_k}
                \quad & \alpha_1\sum_{k\in \mathcal{K}}(\mu_k^{DL}e_k^{DL}-\ln \mu_k^{DL})\\ 
                & +\alpha_2\sum_{l\in \mathcal{L}}(\mu_l^{UL}e_l^{UL}-\ln \mu_l^{UL})\label{eq:25a}
        \end{split}\\
        \textrm{s.t.} \quad & \sum_{k\in \mathcal{K}} {\big| \boldsymbol{w}_k \big|}^2 \le P_{max}^{BS}\label{eq:25b},
    \end{align}
\end{subequations}
since the objective function \eqref{eq:25a} and the constraint \eqref{eq:25b} are convex, the problem $\mathcal{P}_{1.1}$ can be solved by standard methods.
\subsubsection{Optimal Power Transmission at Uplink Users}
For a given set of \big\{ $\boldsymbol{U}_l, u_k, \mu_k^{DL}, \mu_l^{UL},  \boldsymbol{\mathit{w}}_k$  \big\}, we have the following optimization problem $\mathcal{P}_{1.2}$, in order to derive optimal transmission power of users at uplink.
\begin{subequations}
    \begin{align}
        \begin{split}
                \mathcal{P}_{1.2}:\min_{p_l}
                \quad & \alpha_1\sum_{k\in \mathcal{K}}(\mu_k^{DL}e_k^{DL}-\ln \mu_k^{DL})\\ 
                & +\alpha_2\sum_{l\in \mathcal{L}}(\mu_l^{UL}e_l^{UL}-\ln \mu_l^{UL})\label{eq:26a}
        \end{split}\\
        \textrm{s.t.}\quad & {p_l}^2 \le P_{max}^{l}, \forall  l\label{eq:26b}.
    \end{align}
\end{subequations}
Where $p_l=\sqrt{\rho_l}$. Since  the objective function \eqref{eq:26a} and the constraint \eqref{eq:26b} are convex, similar to the optimization problem $\mathcal{P}_{1.1}$, the problem $\mathcal{P}_{1.2}$ is convex. 
\subsubsection{Optimal Values of \big\{ $\boldsymbol{U}_l, u_k, \mu_k^{DL}, \mu_l^{UL}$ \big\}  } For a given set of \big\{ $u_k, \mu_k^{DL}, \mu_l^{UL}, \boldsymbol{\mathit{w}}_k, \rho_l$  \big\}, we first transform problem $\mathcal{P}_{1}$ into problem $\mathcal{P}_{1.3}$ to find optimal value of $\boldsymbol{U}_l$ as follows:
\\
\begin{equation}
    \mathcal{P}_{1.3}:\quad \min_{\boldsymbol{U}_l}
    \quad  \alpha_2\sum_{l\in \mathcal{L}}(\mu_l^{UL}e_l^{UL}-\ln \mu_l^{UL})\label{eq:27}.
\end{equation}
Since, the objective function \eqref{eq:27} is a convex function of $\boldsymbol{U}_l$, by taking the first derivative of \eqref{eq:27} with respect to $\boldsymbol{U}_l$ and set it equal to zero, the optimal value of  the combining vector at the BS is derived as
\begin{equation}
\begin{split}
    \boldsymbol{U}_1^{opt}&=\Bigg({\big|\boldsymbol{H}_{r3}\sqrt{\rho_1}\big|}^2 +{\big|\boldsymbol{G}_{t3}\sqrt{\rho_2}\big|}^2 + \bigg(\hat{\sigma}^2{\big|(\boldsymbol{\mathit{w}}_1 + \boldsymbol{\mathit{w}}_2)\big|}^2\\
    &+\sigma_{UL}^2\bigg)\boldsymbol{\mathrm{I}}_{N_t} \Bigg)^{-1}\boldsymbol{H}_{r3}\sqrt{\rho_1},\label{eq:28}
\end{split}
\raisetag{17pt}
\end{equation}
and
\begin{equation}
    \begin{split}
    \boldsymbol{U}_2^{opt}&=\Bigg({\big|\boldsymbol{H}_{r3}\sqrt{\rho_1}\big|}^2 + {\big|\boldsymbol{G}_{t3}\sqrt{\rho_2}\big|}^2 + \bigg(\hat{\sigma}^2{\big|(\boldsymbol{\mathit{w}}_1 + \boldsymbol{\mathit{w}}_2)\big|}^2\\
    & +\sigma_{UL}^2\bigg)\boldsymbol{\mathrm{I}}_{N_t} \Bigg)^{-1}\boldsymbol{G}_{t3}\sqrt{\rho_2}.\label{eq:29}
    \end{split}
    \raisetag{17pt}
\end{equation}
Similarly, for a given set of \big\{ $\boldsymbol{U}_l, \mu_k^{DL}, \mu_l^{UL}, \boldsymbol{\mathit{w}}_k, \rho_l$  \big\}, the problem $\mathcal{P}_1$ is transformed into
\\
\begin{equation}
    \mathcal{P}_{1.4}:\quad \min_{u_k}
    \quad  \alpha_1\sum_{k\in \mathcal{K}}(\mu_k^{DL}e_k^{DL}-\ln \mu_k^{DL})\label{eq:30}.
\end{equation}
 Therefore, in order to derive the optimal value of the decoding coefficient, we compute the first derivative of \eqref{eq:30} with respect to $u_k$ and set it to zero, we have 
 \\
 \begin{equation}
    u_k^{opt}=\frac{\boldsymbol{H}_{rk}\boldsymbol{\mathit{w}}_k}{{\big|\boldsymbol{H}_{rk}\boldsymbol{\mathit{w}}_k \big|}^2 + \big|\boldsymbol{G}_{tk}\sqrt{\rho_{\acute{k}}}\big|^2 + {\big|\boldsymbol{H}_{rk}\boldsymbol{\mathit{w}}_{\acute{k}} \big|}^2 + \sigma_{DL}^2}. \label{eq:31} 
\end{equation}
\\
Eventually, to gain the optimal values of $\mu_k^{DL}$ and $\mu_l^{UL}$, since the objective function of \eqref{eq:21a} is convex with respect to $\mu_k^{DL}$ and $\mu_l^{UL}$, by taking first derivative of \eqref{eq:21a} with respect to  $\mu_k^{DL}$ and $\mu_l^{UL}$ separately and then set them to zero, we have 
\\
\begin{equation}
    {\mu_k^{DL}}^{opt}={e_k^{DL}}^{-1},\label{eq:32}
\end{equation}
\begin{equation}
    {\mu_l^{UL}}^{opt}={e_l^{UL}}^{-1}.\label{eq:33}
\end{equation}\vfill
According to the above analyses the alternating procedure to solve the sub-problems of $\mathcal{P}_{1}$ is summarized in \textbf{Algorithm 1}.
\begin{algorithm}
\caption{: Iterative Algorithm to Solve $\mathcal{P}_{1}$ Given in \eqref{P4}}\label{alg:cap1}
\textbf{Input}: {Channel coefficients $\boldsymbol{H}_{r1}, \boldsymbol{G}_{t1}, \boldsymbol{H}_{r2}, \boldsymbol{G}_{t2}, \boldsymbol{H}_{r3}, \boldsymbol{G}_{t3}$}. Maximum powers $P_{max}^{BS},P_{max}^{l}, \forall l$. Initial values for $p_l^{(0)}, {\boldsymbol{\mathit{w}}_k}^{(0)}$ and stopping accuracy $\epsilon_1$.
\begin{algorithmic}[1]
\For{$n=1, 2, \dots$}
    \State Update ${\boldsymbol{\mathit{w}}_k}^{(n)}$ by solving $\mathcal{P}_{1.1}$.
    \State Update ${\rho_l}^{(n)}$ by solving $\mathcal{P}_{1.2}$.
    \State Update ${\boldsymbol{U}_1}^{(n)}$  using \eqref{eq:28}.
    \State Update ${\boldsymbol{U}_2}^{(n)}$  using \eqref{eq:29}. 
    \State Update ${u_k}^{(n)}$ using \eqref{eq:31}.
    \State Update $\mu_k^{DL}$ using \eqref{eq:32} while $e_k^{DL}$ is computed as in \eqref{eq:22}. 
    \State Update $\mu_l^{UL}$ using \eqref{eq:33} while $e_l^{UL}$ is computed as in \eqref{eq:23} and \eqref{eq:24}.
    \State \textbf{Until} $\Big|{\textrm{WSR}}^{(n)} - {\textrm{WSR}}^{(n-1)}  \Big| < \epsilon_1 $.
\EndFor
\end{algorithmic}
\textbf{Output}: The optimal solutions: ${\boldsymbol{\mathit{w}}_k}^{opt}={\boldsymbol{\mathit{w}}_k}^{(n)},\forall k,{\rho_l}^{opt}=({p_l}^{(n)})^2,\forall l ,{\boldsymbol{U}_l}^{opt}={\boldsymbol{U}_l}^{(n)},\forall l$ and ${u_k}^{opt}={u_k}^{(n)},\forall k$. 
\end{algorithm}
\subsection{Optimizing $\boldsymbol{\Phi}_t$ and $\boldsymbol{\Phi}_r$ using SCA}\vfill
In this subsection, a similar iterative approach as in \cite{perera2022sum} is employed to solve \eqref{P1}, \eqref{P2} and \eqref{P3}. First, we reformulate the rates of DL/UL users by setting $\boldsymbol{q}^t= {\big[\sqrt{\beta_1^t}e^{j\phi_1^{t}}, \dots, \sqrt{\beta_M^t}e^{j\phi_M^{t}} \big]}^T$ and $\boldsymbol{q}^r= {\big[\sqrt{\beta_1^r}e^{j\phi_1^{r}}, \dots, \sqrt{\beta_M^r}e^{j\phi_M^{r}} \big]}^T$. Note that the TARCs are $\boldsymbol{\Phi}_t=\textrm{diag}(\boldsymbol{q}^t)$ and $\boldsymbol{\Phi}_r=\textrm{diag}(\boldsymbol{q}^r)$. For the sake of simplicity, we set $(\boldsymbol{q}^r)^T\boldsymbol{h}_1=\boldsymbol{v}_d^H\boldsymbol{\Phi}_r\boldsymbol{h}_d$, $(\boldsymbol{q}^t)^T\boldsymbol{h}_2=\boldsymbol{v}_d^H\boldsymbol{\Phi}_t\boldsymbol{g}_u$, $(\boldsymbol{q}^t)^T\boldsymbol{h}_3=\boldsymbol{g}_d^H\boldsymbol{\Phi}_t\boldsymbol{h}_d$, $(\boldsymbol{q}^t)^T\boldsymbol{h}_4=\boldsymbol{g}_d^H\boldsymbol{\Phi}_t\boldsymbol{v}_u$, $\boldsymbol{h}_5\boldsymbol{q}^r=\boldsymbol{h}_u^H\boldsymbol{\Phi}_r\boldsymbol{v}_u$ and $\boldsymbol{h}_6\boldsymbol{q}^t=\boldsymbol{h}_u^H\boldsymbol{\Phi}_t\boldsymbol{g}_u$, where $\boldsymbol{h}_1=\textrm{diag}(\boldsymbol{v}_d^H)\boldsymbol{h}_d$, $\boldsymbol{h}_2=\textrm{diag}(\boldsymbol{v}_d^H)\boldsymbol{g}_u$, $\boldsymbol{h}_3=\textrm{diag}(\boldsymbol{g}_d^H)\boldsymbol{h}_d$, $\boldsymbol{h}_4=\textrm{diag}(\boldsymbol{g}_d^H)\boldsymbol{v}_u$, $\boldsymbol{h}_5=\boldsymbol{h}_u^H\textrm{diag}(\boldsymbol{v}_u)$ and $\boldsymbol{h}_6=\boldsymbol{h}_u^H\textrm{diag}(\boldsymbol{g}_u)$. Therefore, the SINR of DL users for ES and MS schemes are given as\vfill 
\begin{equation}
    \gamma_1^{DL}=\frac{{\big|((\boldsymbol{q}^r)^T\boldsymbol{h}_1 + \boldsymbol{f}_1 )\boldsymbol{\mathit{w}}_1 \big|}^2}{{\big|((\boldsymbol{q}^t)^T\boldsymbol{h}_2+{f}_3)\sqrt{\rho_2} \big|}^2+{\big|((\boldsymbol{q}^r)^T\boldsymbol{h}_1+ \boldsymbol{f}_1)\boldsymbol{\mathit{w}}_2 \big|}^2 + \sigma_{U_1}^2}\label{eq:34},
 \end{equation}
and\vfill
 \begin{equation}
    \gamma_2^{DL}=\frac{{\big|((\boldsymbol{q}^t)^T\boldsymbol{h}_3 + \boldsymbol{f}_2 )\boldsymbol{\mathit{w}}_2 \big|}^2}{{\big|((\boldsymbol{q}^t)^T\boldsymbol{h}_4+{f}_3)\sqrt{\rho_2} \big|}^2+{\big|((\boldsymbol{q}^t)^T\boldsymbol{h}_3+ \boldsymbol{f}_2)\boldsymbol{\mathit{w}}_1 \big|}^2 + \sigma_{U_2}^2}\label{eq:35}.
 \end{equation}
 Using \eqref{eq:34} and \eqref{eq:35}, corresponding rates of the DL users can be derived as
 \begin{equation}
 R_1^{DL}=\log_2(1+\gamma_1^{DL}),   
 \end{equation}
 and 
 \begin{equation}
 R_2^{DL}=\log_2(1+\gamma_2^{DL}).   
 \end{equation}\vfill
 In addition, the SINR of UL users for ES and MS schemes can be derived as
 \begin{equation}
    \gamma_1^{UL}=\frac{{\big|\boldsymbol{U}_1^H(\boldsymbol{h}_5\boldsymbol{q}^r+\boldsymbol{f}_1^H)\sqrt{\rho_1} \big|}^2}{{\big|\boldsymbol{U}_1^H(\boldsymbol{h}_6\boldsymbol{q}^t+\boldsymbol{f}_2^H)\sqrt{\rho_2} \big|}^2+\boldsymbol{\textrm{RSI}}(\boldsymbol{U}_1) + {\big|\boldsymbol{U}_1^H\big|}^2\sigma_{UL}^2}\label{eq:38},
 \end{equation}
and 
\begin{equation}
    \gamma_2^{UL}=\frac{{\big|\boldsymbol{U}_2^H(\boldsymbol{h}_6\boldsymbol{q}^t+\boldsymbol{f}_2^H)\sqrt{\rho_2} \big|}^2}{{\big|(\boldsymbol{U}_2^H(\boldsymbol{h}_5\boldsymbol{q}^r+\boldsymbol{f}_1^H)\sqrt{\rho_1} \big|}^2+\boldsymbol{\textrm{RSI}}(\boldsymbol{U}_2) + {\big|\boldsymbol{U}_2^H\big|}^2\sigma_{UL}^2}\label{eq:39}.
 \end{equation}
 By using \eqref{eq:38} and \eqref{eq:39}, corresponding rates of the UL users can be defined as
 
 \begin{equation}
 R_1^{UL}=\log_2(1+\gamma_1^{UL}),   
 \end{equation}
 and 
 \begin{equation}
 R_2^{UL}=\log_2(1+\gamma_2^{UL}).   
 \end{equation}

 Note that in the TS scheme, the interference terms in the denominator of SINR is eliminated.\\
 Then, methodical algorithms are invoked to design STAR-RIS coefficients of the three aforesaid protocols by considering the following WSR problem as follows: 
 \\
\begin{subequations}
\label{P5}
\begin{align}
\mathcal{P}_2:\max_{\boldsymbol{q}^w,\forall w}
\quad & \alpha_1\sum_{k\in \mathcal{K}}R_k^{DL}+\alpha_2\sum_{l\in \mathcal{L}}R_l^{UL}\\
\textrm{s.t.}
 \quad & {\big|q_{m,x}^t \big|}^2 + {\big|q_{m,x}^r \big|}^2 = 1, \forall m \in \mathcal{M}, \label{eq:42b}    \\
 \quad & 0 \le {\big|q_{m,x}^w\big|} \le 1, \forall m \in \mathcal{M}, \textrm{for ES}, \label{eq:42c} \\ 
 \quad & {\big|q_{m,x}^w\big|} \in \big\{0,1 \big\}, \forall m \in \mathcal{M}, \textrm{for MS}, \label{eq:42d}
\end{align}
\end{subequations}
\\
where $w \in \big\{t,r \big\}$ and $x \in \big\{\textrm{ES,MS} \big\}$. The constraint \eqref{eq:42b}  models  amplitude limitation of STAR-RIS, (i.e. \eqref{eq:18f} and \eqref{eq:19f}). The constraints \eqref{eq:42c} and \eqref{eq:42d} are amplitude constraints  for the ES and MS protocols, (i.e. \eqref{eq:18e} and \eqref{eq:19e}), respectively.\\
Next, we apply the SCA technique to cope with the non-concavity of the UL/DL user rates.\\
In particular, according to \cite{niu2021simultaneous,nniu2021weighted,sheng2018beamforming}, around given point $\Big\{\tilde{x}, \tilde{y} \Big\}$, the following inequality holds:
\\
\begin{equation}
    \begin{split}
        &\ln{\Bigg(1+\frac{{\big|x\big|}^2}{y}\Bigg)} \ge \ln{\Bigg(1+\frac{{\big|\tilde{x}\big|}^2}{\tilde{y}}\Bigg)} - \frac{{\big|\tilde{x}\big|}^2}{\tilde{y}} + \frac{2\mathfrak{R}\big\{x\tilde{x}\big\}}{\tilde{y}}\\
        &-\frac{{\big|\tilde{x}\big|}^2\Big( y + {\big|x\big|}^2 \Big)}{\tilde{y}\Big( \tilde{y} + {\big|\tilde{x}\big|}^2 \Big)}.
    \end{split}
\end{equation}
\\
Therefore, the following lower bounds for $R_1^{DL}, R_2^{DL}, R_1^{UL}$  and $R_2^{UL} $ are derived as:
\\
\begin{equation}
    \begin{split}
        &R_1^{DL} \ge R_{1,i}^{DL,lb} , R_2^{DL} \ge R_{2,i}^{DL,lb} , R_1^{UL} \ge R_{1,i}^{UL,lb}\\
        &\textrm{and} \quad  R_2^{UL} \ge R_{2,i}^{UL,lb},
    \end{split}
\end{equation}\\
where $R_{1,i}^{DL,lb}, R_{2,i}^{DL,lb}, R_{1,i}^{UL,lb} $ and $R_{2,i}^{UL,lb}$ are given as follows:
\\
\begin{equation}
\begin{split}
\label{eq:45} R_{1,i}^{DL,lb} &= \frac{1}{\ln(2)}\Bigg[\ln\Bigg(1+ \frac{{\big|\chi_1^i\big|}^2}{\zeta_1^i}\Bigg)- \frac{{\big|\chi_1^i\big|}^2}{\zeta_1^i}\\
&+ \frac{2\mathfrak{R}\big\{(\chi_1^i)^H\chi_1 \big\}}{\zeta_1^i}
-\frac{{\big|\chi_1^i\big|}^2}{\zeta_1^i(\zeta_1^i+{\big|\chi_1^i\big|}^2)}\Bigg( {\big|\chi_1\big|}^2 + \zeta_1  \Bigg)  \Bigg],
\end{split}
\end{equation}
\begin{equation}
\begin{split}
\label{eq:46} R_{2,i}^{DL,lb} &= \frac{1}{\ln(2)}\Bigg[\ln\Bigg(1+ \frac{{\big|\chi_2^i\big|}^2}{\zeta_2^i}\Bigg)- \frac{{\big|\chi_2^i\big|}^2}{\zeta_2^i}\\
&+ \frac{2\mathfrak{R}\big\{(\chi_2^i)^H\chi_2 \big\}}{\zeta_2^i}
-\frac{{\big|\chi_2^i\big|}^2}{\zeta_2^i(\zeta_2^i+{\big|\chi_2^i\big|}^2)}\Bigg( {\big|\chi_2\big|}^2 + \zeta_2  \Bigg)  \Bigg],
\end{split}
\end{equation}
\begin{equation}
\begin{split}
\label{eq:47} R_{1,i}^{UL,lb} &= \frac{1}{\ln(2)}\Bigg[\ln\Bigg(1+ \frac{{\big|\chi_3^i\big|}^2}{\zeta_3^i}\Bigg)- \frac{{\big|\chi_3^i\big|}^2}{\zeta_3^i}\\
&+ \frac{2\mathfrak{R}\big\{(\chi_3^i)^H\chi_3 \big\}}{\zeta_3^i}
-\frac{{\big|\chi_3^i\big|}^2}{\zeta_3^i(\zeta_3^i+{\big|\chi_3^i\big|}^2)}\Bigg( {\big|\chi_3\big|}^2 + \zeta_3  \Bigg)  \Bigg],
\end{split}
\end{equation}
\begin{equation}
\begin{split}
\label{eq:48} R_{2,i}^{UL,lb} &= \frac{1}{\ln(2)}\Bigg[\ln\Bigg(1+ \frac{{\big|\chi_4^i\big|}^2}{\zeta_4^i}\Bigg)- \frac{{\big|\chi_4^i\big|}^2}{\zeta_4^i}\\
&+ \frac{2\mathfrak{R}\big\{(\chi_4^i)^H\chi_4 \big\}}{\zeta_4^i}
-\frac{{\big|\chi_4^i\big|}^2}{\zeta_4^i(\zeta_4^i+{\big|\chi_4^i\big|}^2)}\Bigg( {\big|\chi_4\big|}^2 + \zeta_4  \Bigg)  \Bigg],
\end{split}
\end{equation}
where $\chi_1^i=((\boldsymbol{q}_i^r)^T\boldsymbol{h}_1 + \boldsymbol{f}_1 )\boldsymbol{\mathit{w}}_1$, $\zeta_1^i={\big|((\boldsymbol{q}_i^t)^T\boldsymbol{h}_2+{f}_3)\sqrt{\rho_2} \big|}^2+{\big|((\boldsymbol{q}_i^r)^T\boldsymbol{h}_1+ \boldsymbol{f}_1)\boldsymbol{\mathit{w}}_2 \big|}^2 + \sigma_{U_1}^2$, $\chi_2^i=((\boldsymbol{q}_i^t)^T\boldsymbol{h}_3 + \boldsymbol{f}_2 )\boldsymbol{\mathit{w}}_2$, $\zeta_2^i={\big|((\boldsymbol{q}_i^t)^T\boldsymbol{h}_4+{f}_3)\sqrt{\rho_2} \big|}^2+{\big|((\boldsymbol{q}_i^t)^T\boldsymbol{h}_3+ \boldsymbol{f}_2)\boldsymbol{\mathit{w}}_1 \big|}^2 + \sigma_{U_2}^2$, $\chi_3^i=\boldsymbol{U}_1^H(\boldsymbol{h}_5\boldsymbol{q}_i^r+\boldsymbol{f}_1^H)\sqrt{\rho_1}$, $\zeta_3^i={\big|\boldsymbol{U}_1^H(\boldsymbol{h}_6\boldsymbol{q}_i^t+\boldsymbol{f}_2^H)\sqrt{\rho_2} \big|}^2+\boldsymbol{\textrm{RSI}}(\boldsymbol{U}_1) + \sigma_{UL}^2$, $\chi_4^i=\boldsymbol{U}_2^H(\boldsymbol{h}_6\boldsymbol{q}_i^t+\boldsymbol{f}_2^H)\sqrt{\rho_2}$ and $\zeta_4^i={\big|(\boldsymbol{U}_2^H(\boldsymbol{h}_5\boldsymbol{q}_i^r+\boldsymbol{f}_1^H)\sqrt{\rho_1} \big|}^2+\boldsymbol{\textrm{RSI}}(\boldsymbol{U}_2) + \sigma_{UL}^2$. In addition, i denotes the iteration index. It is noteworthy that 
\begin{comment}
\begin{figure*}[!t]
% ensure that we have normalsize text
\normalsize
\begin{equation}
\label{eq:46} R_{2,i}^{DL,lb} = \frac{1}{\ln(2)}\Bigg[\ln\Bigg(1+ \frac{{\big|\chi_2^i\big|}^2}{\zeta_2^i}\Bigg)  + \frac{2\mathfrak{R}\big\{(\chi_2^i)^H\chi_2 \big\}}{\zeta_2^i}-\frac{{\big|\chi_2^i\big|}^2}{\zeta_2^i(\zeta_2^i+{\big|\chi_2^i\big|}^2)}\Bigg( {\big|\chi_2\big|}^2 + \zeta_2  \Bigg) - \frac{{\big|\chi_2^i\big|}^2}{\zeta_2^i} \Bigg]
\end{equation}
\begin{equation}
\label{eq:47} R_{1,i}^{UL,lb} = \frac{1}{\ln(2)}\Bigg[\ln\Bigg(1+ \frac{{\big|\chi_3^i\big|}^2}{\zeta_3^i}\Bigg)  + \frac{2\mathfrak{R}\big\{(\chi_3^i)^H\chi_3 \big\}}{\zeta_3^i}-\frac{{\big|\chi_3^i\big|}^2}{\zeta_3^i(\zeta_3^i+{\big|\chi_3^i\big|}^2)}\Bigg( {\big|\chi_3\big|}^2 + \zeta_3  \Bigg) - \frac{{\big|\chi_3^i\big|}^2}{\zeta_3^i} \Bigg]
\end{equation}
\begin{equation}
\label{eq:48} R_{2,i}^{UL,lb} = \frac{1}{\ln(2)}\Bigg[\ln\Bigg(1+ \frac{{\big|\chi_4^i\big|}^2}{\zeta_4^i}\Bigg)  + \frac{2\mathfrak{R}\big\{(\chi_4^i)^H\chi_4 \big\}}{\zeta_4^i}-\frac{{\big|\chi_4^i\big|}^2}{\zeta_4^i(\zeta_4^i+{\big|\chi_4^i\big|}^2)}\Bigg( {\big|\chi_4\big|}^2 + \zeta_4  \Bigg) - \frac{{\big|\chi_4^i\big|}^2}{\zeta_4^i} \Bigg]
\end{equation}
\hrulefill
\vspace*{4pt}
\end{figure*}
\end{comment}
$R_{1,i}^{DL,lb}, R_{2,i}^{DL,lb}, R_{1,i}^{UL,lb} $ and $R_{2,i}^{UL,lb}$ are concave function of $q_x^t$ and $q_x^r$. This is significant becuase the reformulated problem maximizes a lower bound of the original objective of $\mathcal{P}_{2}$ subject to the constraints. Therefore, the approximation of reformulated problem $\mathcal{P}_{2}$ for the ES and MS protocols is proposed in the following subsections. 
\subsubsection{Proposed Solution for the ES Protocol}
By invoking \eqref{eq:45}, \eqref{eq:46}, \eqref{eq:47} and \eqref{eq:48}, the $i$-th iteration of the problem $\mathcal{P}_2$ can be written as 
\begin{subequations}
\label{P6}
\begin{align}
\mathcal{P}_{2.1}:\max_{\boldsymbol{q}^w , \tilde{\mathcal{R}}_k^{DL}, \tilde{\mathcal{R}}_l^{UL} }
\quad & \alpha_1\sum_{k\in \mathcal{K}}\tilde{\mathcal{R}}_k^{DL}+\alpha_2\sum_{l\in \mathcal{L}}\tilde{\mathcal{R}}_l^{UL}\label{eq:49a}\\
\textrm{s.t.} \quad & \tilde{\mathcal{R}}_k^{DL}\le R_{k,i}^{DL,lb}, \forall k \in \mathcal{K},\label{eq:49b}   \\
\quad & \tilde{\mathcal{R}}_l^{UL}\le R_{l,i}^{UL,lb}, \forall l \in \mathcal{L},  \label{eq:49c} \\
 \quad & {\big|q_{m}^t \big|}^2 + {\big|q_{m}^r \big|}^2 \le 1, \forall m \in \mathcal{M}, \label{eq:49d}    \\
 \quad & 0 \le {\big|q_{m}^w\big|} \le 1, \forall m \in \mathcal{M},  \label{eq:49e} 
\end{align}
\end{subequations}
where $\tilde{\mathcal{R}}_k^{DL}$ and $\tilde{\mathcal{R}}_l^{UL}$ are auxiliary variables. In $\mathcal{P}_{2.1}$, the objective function and all constraints are convex and then can be efficiently solved by the standard convex optimization methods. Note that the equality constraint in \eqref{eq:42b} is relaxed in \eqref{eq:49d} in order to make it convex. At the optimal solution, \eqref{eq:49d} satisfies with equality. According to the last analysis, the iterative scheme to solve the problem \eqref{P1} is summarized in \textbf{Algorithm 2}.   
\begin{algorithm}
\caption{: Iterative  Algorithm for the ES Protocol to Solve Problem Given in \eqref{P1}}\label{alg:cap2}
\textbf{Input}: Initial values for $\boldsymbol{q}_{(0)}^t$ and $\boldsymbol{q}_{(0)}^r$, Channel coefficients $\boldsymbol{H}_{r1}$, $\boldsymbol{G}_{t1}$, $\boldsymbol{H}_{r2}$, $\boldsymbol{G}_{t2}$, $\boldsymbol{H}_{r3}$, $\boldsymbol{G}_{t3}$. Maximum powers $P_{max}^{BS},P_{max}^{l}, \forall l$. Initial values for $p_l^{(0)}, {\boldsymbol{\mathit{w}}_k}^{(0)}$ and tolerances $\big\{\epsilon_1, \epsilon_2, \epsilon_3\big\}$. 
\begin{algorithmic}[1]
\For{$i=1, 2, \dots$}
    \State
    \parbox[t]{\dimexpr\linewidth-\algorithmicindent}{%
    For given ${\boldsymbol{\mathit{w}}_k}^{(i)}, {u_k}^{(i)},\forall k, {\boldsymbol{U}_l}^{(i)}$ and ${\rho_l}^{(i)}, \forall l,   $ update $\boldsymbol{q}_{(i)}^t$ and $\boldsymbol{q}_{(i)}^r$.
    }
    \For{$j=1, 2, \dots$}
    \State
    \parbox[t]{\dimexpr\linewidth-\algorithmicindent}{%
    For given $\boldsymbol{q}_{(j)}^t$ and $\boldsymbol{q}_{(j)}^r$, solve the problem $\mathcal{P}_{2.1}$\\ and obtain the optimal solution ${\boldsymbol{q}^t}^*$ and ${\boldsymbol{q}^r}^*$. 
    }
    \State
    \parbox[t]{\dimexpr\linewidth-\algorithmicindent}{%
    \textbf{Until}: If the fraction increase of objective value\\ of $\mathcal{P}_{2.1}$ is below $\epsilon_2$; obtain $\boldsymbol{q}_{(i+1)}^t=\boldsymbol{q}_{(j+1)}^t$\\ and $\boldsymbol{q}_{(i+1)}^r=\boldsymbol{q}_{(j+1)}^r$.
    }
    \EndFor
    \State
    \parbox[t]{\dimexpr\linewidth-\algorithmicindent}{%
    For given $\boldsymbol{q}_{(i+1)}^t$ and $\boldsymbol{q}_{(i+1)}^r$ update ${\boldsymbol{\mathit{w}}_k}^{(i+1)}$, ${u_k}^{(i+1)}$,$\forall k$,  ${\boldsymbol{U}_l}^{(i+1)}$ and ${\rho_l}^{(i+1)}$, $\forall l$ by solving problem $\mathcal{P}_{1}$ using \textbf{Algorithm1}.
    }
    \State \textbf{Until} $\Big|{\textrm{WSR}}^{(i+1)} - {\textrm{WSR}}^{(i)}  \Big| < \epsilon_3 $.
\EndFor
\end{algorithmic}
\textbf{Output}: The optimal solutions: ${\boldsymbol{\mathit{w}}_k}^{\textrm{opt}}={\boldsymbol{\mathit{w}}_k}^{(i)},\forall k,{\rho_l}^{\textrm{opt}}={\rho_l}^{(i)},\forall l ,{\boldsymbol{U}_l}^{\textrm{opt}}={\boldsymbol{U}_l}^{(i)},\forall l$ and ${u_k}^{\textrm{opt}}={u_k}^{(i)},\forall k$,         $\boldsymbol{q}_{\textrm{opt}}^t=\boldsymbol{q}_{(i+1)}^t$ and $\boldsymbol{q}_{\textrm{opt}}^r=\boldsymbol{q}_{(i+1)}^r$ .
\end{algorithm}
\subsubsection{Proposed Solution for the MS Protocol} Similar to the ES protocol, by using \eqref{eq:45}, \eqref{eq:46}, \eqref{eq:47} and \eqref{eq:48}, the reformulated problem can be written as
\hspace\fill
\begin{subequations}
\label{P7}
\begin{align}
\mathcal{P}_{2.2}:\max_{\boldsymbol{q}^w ,  \tilde{\mathcal{R}}_k^{DL}, \tilde{\mathcal{R}}_l^{UL} }
\quad & \alpha_1\sum_{k\in \mathcal{K}}\tilde{\mathcal{R}}_k^{DL}+\alpha_2\sum_{l\in \mathcal{L}}\tilde{\mathcal{R}}_l^{UL}\\
\textrm{s.t.} \quad & \tilde{\mathcal{R}}_k^{DL}\le R_{k,i}^{DL,lb}, \forall k \in \mathcal{K},\label{eq:50b}   \\
\quad & \tilde{\mathcal{R}}_l^{UL}\le R_{l,i}^{UL,lb}, \forall l \in \mathcal{L},  \label{eq:50c} \\
 \quad & {\big|q_{m}^t \big|}^2 + {\big|q_{m}^r \big|}^2 \le 1, \forall m \in \mathcal{M}, \label{eq:50d}    \\
 \quad &  {\big|q_{m}^w\big|} \in \big\{0,1\big\}, \forall m \in \mathcal{M},  \label{eq:50e} 
\end{align}
\end{subequations}
the main difference is the constraint ${\big|q_{m}^w\big|} \in \big\{0,1\big\}$. Thus, we need to solve a mixed-integer problem, due to binary constraint, which is non-convex and NP-hard in general. To reduce the complexity, we use a penalty-based scheme in which the non-convex constraint is transformed into convex term and added to the objective function.\\
It is noticeable $\varphi_m^k-(\varphi_m^k)^2 \ge 0, \forall k \in \big\{t,r\big\}, m \in \mathcal{M} $ is always satisfied by the binary constraint, where $\varphi_m^k$ is the amplitude of $q_m^k$. In addition, the equality holds if and only if $\varphi_m^k$ is a binary variable. An upper bound can be derived by using Taylor series expansion as
\begin{equation}
    \begin{split}
        \varphi_m^w-(\varphi_m^w)^2 &\le (\varphi_m^{w,i})^2 + (1-2\varphi_m^{w,i})\varphi_m^w,\\
        & = f(\varphi_m^{w,i},\varphi_m^w), \forall m \in \mathcal{M}, 
    \end{split}
\end{equation}\vfill
where $w \in \big\{ t,r\big\}$ and $\varphi_m^{w,i}$ is a given point in the $i$-th iteration of SCA. The problem $\mathcal{P}_{2.2}$ is transformed to $\mathcal{P}_{2.3}$ by adding the upper bound for binary constraint as a penalty term as:\vfill
\begin{subequations}
\label{P8}
\begin{align}
\mathcal{P}_{2.3}:\max_{\substack{\boldsymbol{q}^w , \tilde{\mathcal{R}}_k^{DL}, \tilde{\mathcal{R}}_l^{UL},\\ \varphi_m^w} }
\quad & \alpha_1\sum_{k\in \mathcal{K}}\tilde{\mathcal{R}}_k^{DL}+\alpha_2\sum_{l\in \mathcal{L}}\tilde{\mathcal{R}}_l^{UL} - \mu\Delta \label{eq:52a}\\
\textrm{s.t.} \quad & \eqref{eq:50b},\eqref{eq:50c},\\
 \quad & {(\varphi_m^t )}^2 + {(\varphi_m^r )}^2 \le 1, \forall m \in \mathcal{M}, \label{eq:52c}    \\
 \quad &  {\big|q_{m}^w\big|} \le \varphi_m^w , \forall m \in \mathcal{M},  \label{eq:52d} 
\end{align}
\end{subequations}
where $\Delta=\sum_{m=1}^{\mathcal{M}}\sum_{w \in \left\{t,r\right\}} f(\varphi_m^{w,i},\varphi_m^w)$ , and the constraint \eqref{eq:52d} holds equality at optimum state  and $\mu$ is the penalty factor. The termination condition for satisfying the constraint \eqref{eq:50e} has to be fallen below $\epsilon_3$. Besides, the objective function and constraints of  $\mathcal{P}_{2.3}$ are convex and can be efficiently solved by the standard convex optimization
methods. According to the above analysis, the iterative scheme to solve the problem \eqref{P2} is summarized in \textbf{Algorithm 3}. 
\begin{algorithm}
\caption{: Iterative  Algorithm for the MS Protocol to Solve Problem Given in \eqref{P2}}\label{alg:cap3}
\textbf{Input}: Initial values for $\boldsymbol{q}_{(0)}^t$ and $\boldsymbol{q}_{(0)}^r$, ${\boldsymbol{\varphi}^t}^{(0)}$ and ${\boldsymbol{\varphi}^r}^{(0)}$ ,Channel coefficients $\boldsymbol{H}_{r1}$, $\boldsymbol{G}_{t1}$, $\boldsymbol{H}_{r2}$, $\boldsymbol{G}_{t2}$, $\boldsymbol{H}_{r3}$, $\boldsymbol{G}_{t3}$. Maximum powers $P_{max}^{BS},P_{max}^{l}, \forall l$. Initial values for $p_l^{(0)}, {\boldsymbol{\mathit{w}}_k}^{(0)}$,  tolerances $\big\{\epsilon_1, \epsilon_2, \epsilon_3, \epsilon_4 \big\}$ and the penalty parameters $\mu$, $\omega$. 
\begin{algorithmic}[1]
\For{$i=1, 2, \dots$}
    \State 
    \parbox[t]{\dimexpr\linewidth-\algorithmicindent}{%
    For given ${\boldsymbol{\mathit{w}}_k}^{(i)}, {u_k}^{(i)},\forall k, {\boldsymbol{U}_l}^{(i)}$ and ${\rho_l}^{(i)}, \forall l,   $ update $\boldsymbol{q}_{(i)}^t$ and $\boldsymbol{q}_{(i)}^r$.
    }
    \For{$j=1, 2, \dots$}
    \For{$k=1, 2, \dots$}
    \State
    \parbox[t]{\dimexpr\linewidth-\algorithmicindent}{%
    For given $\boldsymbol{q}_{(k)}^t$ and $\boldsymbol{q}_{(k)}^r$, solve the problem\\ $\mathcal{P}_{2.3}$ and obtain the optimal solution ${\boldsymbol{q}^t}^*$  and\\ ${\boldsymbol{q}^r}^*.$
    }
    \State
    \parbox[t]{\dimexpr\linewidth-\algorithmicindent}{%
    \textbf{Until}: If the fraction increase of objective\\ value of $\mathcal{P}_{2.3}$ is below $\epsilon_2$.   
    }
    \EndFor
    \State
    \parbox[t]{\dimexpr\linewidth-\algorithmicindent}{%
    Obtain $\boldsymbol{q}_{(i+1)}^t=\boldsymbol{q}_{(j+1)}^t$, $\boldsymbol{q}_{(i+1)}^r=\boldsymbol{q}_{(j+1)}^r$ and\\ $\mu=\omega\mu$.
    }
    \State
    \parbox[t]{\dimexpr\linewidth-\algorithmicindent}{%
    \textbf{Until}:$\max\left\{\big|\varphi_m^w\big|-{\big|\varphi_m^w\big|}^2,\forall w \in \left\{t,r\right\}, m \in \mathcal{M}  \right\}$\\  $  \le \epsilon_3 $ 
    }
    \EndFor
    
    \State
    \parbox[t]{\dimexpr\linewidth-\algorithmicindent}{%
    For given $\boldsymbol{q}_{(i+1)}^t$ and $\boldsymbol{q}_{(i+1)}^r$ update ${\boldsymbol{\mathit{w}}_k}^{(i+1)}$, ${u_k}^{(i+1)}$,$\forall k$,  ${\boldsymbol{U}_l}^{(i+1)}$ and ${\rho_l}^{(i+1)}$, $\forall l$ by solving problem $\mathcal{P}_{1}$ using \textbf{Algorithm1}.
    }
    \State \textbf{Until} $\Big|{\textrm{WSR}}^{(i+1)} - {\textrm{WSR}}^{(i)}  \Big| < \epsilon_4 $.
\EndFor
\end{algorithmic}
\textbf{Output}: The optimal solutions: ${\boldsymbol{\mathit{w}}_k}^{\textrm{opt}}={\boldsymbol{\mathit{w}}_k}^{(i)},\forall k,{\rho_l}^{\textrm{opt}}={\rho_l}^{(i)},\forall l ,{\boldsymbol{U}_l}^{\textrm{opt}}={\boldsymbol{U}_l}^{(i)},\forall l$ and ${u_k}^{\textrm{opt}}={u_k}^{(i)},\forall k$,         $\boldsymbol{q}_{\textrm{opt}}^t=\boldsymbol{q}_{(i+1)}^t$ and $\boldsymbol{q}_{\textrm{opt}}^r=\boldsymbol{q}_{(i+1)}^r$ .
\end{algorithm}
\subsubsection{Proposed Solution for the TS Protocol } For the TS protocol, due to the interference-free feature of the TS protocol, the rates of corresponding users are given as
\begin{equation}
R_1^{DL}=\tau_1^{DL}\log_2\bigg(1+\frac{{\big|((\boldsymbol{q}^r)^T\boldsymbol{h}_1 + \boldsymbol{f}_1 )\boldsymbol{\mathit{w}}_1 \big|}^2}{ \sigma_{DL}^2}\bigg)\label{eq:53},\hspace{0.5cm}    
\end{equation}\vfill
\begin{equation}
R_2^{DL}=\tau_2^{DL}\log_2\bigg(1+\frac{{\big|((\boldsymbol{q}^t)^T\boldsymbol{h}_3 + \boldsymbol{f}_2 )\boldsymbol{\mathit{w}}_2 \big|}^2}{ \sigma_{DL}^2}\bigg)\label{eq:54},
\hspace{0.5cm}
\end{equation}
\begin{equation}
R_1^{UL}=\tau_1^{UL}\log_2\bigg(1+\frac{{\big|\boldsymbol{U}_1^H(\boldsymbol{h}_5\boldsymbol{q}^r+\boldsymbol{f}_1^H)\sqrt{\rho_1} \big|}^2}{ {\big|\boldsymbol{U}_1^H\big|}^2\sigma_{UL}^2}\bigg)\label{eq:55},    
\end{equation}
\begin{equation}
R_2^{UL}=\tau_2^{UL}\log_2\bigg(1+\frac{{\big|\boldsymbol{U}_2^H(\boldsymbol{h}_6\boldsymbol{q}^t+\boldsymbol{f}_2^H)\sqrt{\rho_2} \big|}^2}{ {\big|\boldsymbol{U}_2^H\big|}^2\sigma_{UL}^2}\bigg)\label{eq:56}.   
\end{equation}
Thus, the lower bounds for $R_1^{DL}, R_2^{DL}, R_1^{UL}$  and $R_2^{UL} $ for the TS protocol are derived in the following:
\begin{equation}
    \begin{split}
        &R_1^{DL} \ge \check{R}_{1,i}^{DL,lb} , R_2^{DL} \ge \check{R}_{2,i}^{DL,lb} , R_1^{UL} \ge \check{R}_{1,i}^{UL,lb}\\
        &\textrm{and} \quad  R_2^{UL} \ge \check{R}_{2,i}^{UL,lb},
    \end{split}
\end{equation}
where $\check{R}_{1,i}^{DL,lb}, \check{R}_{2,i}^{DL,lb}, \check{R}_{1,i}^{UL,lb} $ and $\check{R}_{2,i}^{UL,lb}$ are given as:
\begin{equation}
\begin{split}
\label{eq:58} \check{R}_{1,i}^{DL,lb} &= \tau_1^{DL}\frac{1}{\ln(2)}\Bigg[\ln\Bigg(1+ \frac{{\big|\psi_1^i\big|}^2}{\kappa_1}\Bigg)- \frac{{\big|\psi_1^i\big|}^2}{\kappa_1}\\
&+ \frac{2\mathfrak{R}\big\{(\psi_1^i)^H\psi_1 \big\}}{\kappa_1}
-\frac{{\big|\psi_1^i\big|}^2}{\kappa_1(\kappa_1+{\big|\psi_1^i\big|}^2)}\Bigg( {\big|\psi_1\big|}^2 + \kappa_1  \Bigg)  \Bigg],
\end{split}
\end{equation}
\begin{equation}
\begin{split}
\label{eq:59} \check{R}_{2,i}^{DL,lb} &= \tau_2^{DL}\frac{1}{\ln(2)}\Bigg[\ln\Bigg(1+ \frac{{\big|\psi_2^i\big|}^2}{\kappa_1}\Bigg)- \frac{{\big|\psi_2^i\big|}^2}{\kappa_1}\\
&+ \frac{2\mathfrak{R}\big\{(\psi_2^i)^H\psi_2 \big\}}{\kappa_1}
-\frac{{\big|\psi_2^i\big|}^2}{\kappa_1(\kappa_1+{\big|\psi_2^i\big|}^2)}\Bigg( {\big|\psi_2\big|}^2 + \kappa_1  \Bigg)  \Bigg],
\end{split}
\end{equation}
\begin{equation}
\begin{split}
\label{eq:60} \check{R}_{1,i}^{UL,lb} &= \tau_1^{UL}\frac{1}{\ln(2)}\Bigg[\ln\Bigg(1+ \frac{{\big|\psi_3^i\big|}^2}{\kappa_2}\Bigg)- \frac{{\big|\psi_3^i\big|}^2}{\kappa_2}\\
&+ \frac{2\mathfrak{R}\big\{(\psi_3^i)^H\psi_3 \big\}}{\kappa_2}
-\frac{{\big|\psi_3^i\big|}^2}{\kappa_2(\kappa_2+{\big|\psi_3^i\big|}^2)}\Bigg( {\big|\psi_3\big|}^2 + \kappa_2  \Bigg)  \Bigg],
\end{split}
\end{equation}
\begin{equation}
\begin{split}
\label{eq:61} \check{R}_{2,i}^{UL,lb} &= \tau_2^{UL}\frac{1}{\ln(2)}\Bigg[\ln\Bigg(1+ \frac{{\big|\psi_4^i\big|}^2}{\kappa_3}\Bigg)- \frac{{\big|\psi_4^i\big|}^2}{\kappa_3}\\
&+ \frac{2\mathfrak{R}\big\{(\psi_4^i)^H\psi_4 \big\}}{\kappa_3}
-\frac{{\big|\psi_4^i\big|}^2}{\kappa_3(\kappa_3+{\big|\psi_4^i\big|}^2)}\Bigg( {\big|\psi_4\big|}^2 + \kappa_3  \Bigg)  \Bigg],
\end{split}
\end{equation}
where $\psi_1^i=((\boldsymbol{q}_i^r)^T\boldsymbol{h}_1 + \boldsymbol{f}_1 )\boldsymbol{\mathit{w}}_1$, $\kappa_1=\sigma_{DL}^2$, $\psi_2^i=((\boldsymbol{q}_i^t)^T\boldsymbol{h}_3 + \boldsymbol{f}_2 )\boldsymbol{\mathit{w}}_2$, $\psi_3^i=\boldsymbol{U}_1^H(\boldsymbol{h}_5\boldsymbol{q}_i^r+\boldsymbol{f}_1^H)\sqrt{\rho_1}$, $\kappa_2={\big|\boldsymbol{U}_1^H\big|}^2\sigma_{UL}^2$, $\psi_4^i=\boldsymbol{U}_2^H(\boldsymbol{h}_6\boldsymbol{q}_i^t+\boldsymbol{f}_2^H)\sqrt{\rho_2}$ and $\kappa_3={\big|\boldsymbol{U}_2^H\big|}^2\sigma_{UL}^2$. 
By using \eqref{eq:58}, \eqref{eq:59}, \eqref{eq:60} and \eqref{eq:61}, the $i$-th iteration of reformulated problem of \eqref{P3} can be defined as
\begin{subequations}
\label{P9}
\begin{align}
\begin{split}
\mathcal{P}_{3}:\max_{\boldsymbol{q}^w , \tilde{\mathcal{R}}_k^{DL}, \tilde{\mathcal{R}}_l^{UL} }
\quad & \alpha_1\sum_{k\in \mathcal{K}}\tilde{\mathcal{R}}_k^{DL} +\alpha_2\sum_{l\in \mathcal{L}}\tilde{\mathcal{R}}_l^{UL}\label{eq:62a}
\end{split}\\
\textrm{s.t.} \quad & \tilde{\mathcal{R}}_k^{DL}\le \check{R}_{k,i}^{DL,lb}, \forall k \in \mathcal{K},\label{eq:62b}  \\
\quad & \tilde{\mathcal{R}}_l^{UL}\le \check{R}_{l,i}^{UL,lb}, \forall l \in \mathcal{L},  \label{eq:62c} \\
 \quad & {\big|q_{m}^w\big|} \le 1, \forall m \in \mathcal{M}, \label{eq:62d} 
\end{align}
\end{subequations}
after obtaining the optimal $\big\{\boldsymbol{q}^t,\boldsymbol{q}^r, \tilde{\mathcal{R}}_{k}^{DL}, \tilde{\mathcal{R}}_{l}^{UL} \big\}$ for a given $\tau_k^{DL}$ and $\tau_l^{UL}$, the one-dimensional search method is used to obtain the optimal $\tau_k^{DL^*}$, $\tau_l^{UL^*}$  and $\big\{\boldsymbol{q}^{t^*},\boldsymbol{q}^{r^*}, \tilde{\mathcal{R}}_k^{DL^*}, \tilde{\mathcal{R}}_l^{UL^*}\big\}$. It can be found that \eqref{eq:62a} is simpler than \eqref{eq:49a} whereas the TS protocol is interference-free scheme. Considering the above analyses the iterative scheme to solve the problem \eqref{P3} is summarized in \textbf{Algorithm 4}. 
\begin{algorithm}
\caption{: Iterative  Algorithm for the TS Protocol to Solve Problem Given in \eqref{P3}}\label{alg:cap4}
\textbf{Input}: Initial values for $\boldsymbol{q}_{(0)}^t$ and $\boldsymbol{q}_{(0)}^r$, Channel coefficients $\boldsymbol{H}_{r1}$, $\boldsymbol{G}_{t1}$, $\boldsymbol{H}_{r2}$, $\boldsymbol{G}_{t2}$, $\boldsymbol{H}_{r3}$, $\boldsymbol{G}_{t3}$. Maximum powers $P_{max}^{BS},P_{max}^{l}, \forall l$. Initial values for $p_l^{(0)}, {\boldsymbol{\mathit{w}}_k}^{(0)}$, $\tau_k^{DL}$, $\tau_l^{UL}$ and tolerances $\big\{\epsilon_1, \epsilon_2, \epsilon_3\big\}$. 
\begin{algorithmic}[1]
\For{$i=1, 2, \dots$}
    \State
    \parbox[t]{\dimexpr\linewidth-\algorithmicindent}{%
    For given ${\boldsymbol{\mathit{w}}_k}^{(i)}, {u_k}^{(i)},\forall k, {\boldsymbol{U}_l}^{(i)}$ and ${\rho_l}^{(i)}, \forall l,   $ update $\boldsymbol{q}_{(i)}^t$ and $\boldsymbol{q}_{(i)}^r$.
    }
    \For{$j=1, 2, \dots$}
    \State
    \parbox[t]{\dimexpr\linewidth-\algorithmicindent}{%
    For given $\boldsymbol{q}_{(j)}^t$ and $\boldsymbol{q}_{(j)}^r$, solve the problem $\mathcal{P}_{3}$\\ and obtain the optimal solution ${\boldsymbol{q}^t}^*$ and ${\boldsymbol{q}^r}^*$.\\ Using
     the one-dimensional search method 
to\\ obtain the optimal ${\tau_k^{DL}}^*$ and ${\tau_l^{UL}}^*$ 
    }
    \State
    \parbox[t]{\dimexpr\linewidth-\algorithmicindent}{%
    \textbf{Until}: If the fraction increase of objective value\\ of $\mathcal{P}_{3}$ is below $\epsilon_2$; obtain $\boldsymbol{q}_{(i+1)}^t=\boldsymbol{q}_{(j+1)}^t$,\\ $\boldsymbol{q}_{(i+1)}^r=\boldsymbol{q}_{(j+1)}^r$, ${\tau_k^{DL}}^{(i+1)}={\tau_k^{DL}}^{(j+1)}$\\ and ${\tau_l^{UL}}^{(i+1)}={\tau_l^{UL}}^{(j+1)}$ .
    }
    \EndFor
    \State
    \parbox[t]{\dimexpr\linewidth-\algorithmicindent}{%
    For given $\boldsymbol{q}_{(i+1)}^t$, $\boldsymbol{q}_{(i+1)}^r$, ${\tau_k^{DL}}^{(i+1)}$ and ${\tau_l^{UL}}^{(i+1)}$  update ${\boldsymbol{\mathit{w}}_k}^{(i+1)}$, ${u_k}^{(i+1)}$,$\forall k$,  ${\boldsymbol{U}_l}^{(i+1)}$ and ${\rho_l}^{(i+1)}$, $\forall l$ by solving problem $\mathcal{P}_{1}$ using \textbf{Algorithm1}.
    }
    \State \textbf{Until} $\Big|{\textrm{WSR}}^{(i+1)} - {\textrm{WSR}}^{(i)}  \Big| < \epsilon_3 $.
\EndFor
\end{algorithmic}
\textbf{Output}: The optimal solutions: ${\boldsymbol{\mathit{w}}_k}^{\textrm{opt}}={\boldsymbol{\mathit{w}}_k}^{(i)},\forall k,{\rho_l}^{\textrm{opt}}={\rho_l}^{(i)},\forall l ,{\boldsymbol{U}_l}^{\textrm{opt}}={\boldsymbol{U}_l}^{(i)},\forall l$ and ${u_k}^{\textrm{opt}}={u_k}^{(i)},\forall k$,         $\boldsymbol{q}_{\textrm{opt}}^t=\boldsymbol{q}_{(i+1)}^t$, $\boldsymbol{q}_{\textrm{opt}}^r=\boldsymbol{q}_{(i+1)}^r$, ${\tau_k^{DL}}^{\textrm{opt}}={\tau_k^{DL}}^{(i+1)}$ and ${\tau_l^{UL}}^{\textrm{opt}}={\tau_l^{UL}}^{(i+1)}$ .
\end{algorithm}
\subsection{Complexity Analysis}
In the precedent sections, we surveyed the maximization of the WSR, i.e., the optimization problems \eqref{P1}, \eqref{P2} and \eqref{P3}, by transforming each the original optimization problem into several sub-problems. The entire iterative approach for solving problems \eqref{P1}, \eqref{P2} and \eqref{P3} is summarized in \textbf{Algorithm 2}, \textbf{Algorithm 3} and \textbf{Algorithm 4}, respectively. In \textbf{Algorithm 1}, the complexity of computing $\boldsymbol{w}_k$ in step 2 is $\mathcal{O}\big({N_t}^3\big)$, the complexity of computing $\boldsymbol{U}_l$ in step 4 and step 5 is $\mathcal{O}\big({N_t}^3\big)$. The iterative WMMSE approach developed in \textbf{Algorithm 1} is based on the block coordinate descent method and its convergence is guaranteed as discussed in \cite{shi2011iteratively}. The proposed iterative algorithms solve a series of convex optimization problems. Specifically, the problem $\mathcal{P}_{2.1}$ in \eqref{eq:49a}-\eqref{eq:49e}, the problem $\mathcal{P}_{2.3}$ in \eqref{eq:52a}-\eqref{eq:52d} and the problem $\mathcal{P}_{3}$ in \eqref{eq:62a}-\eqref{eq:62d} are quadratic-constrained quadratic programming  problems, which are forms of convex optimization problems \cite{boyd2004convex}. These convex optimization problems can be efficiently solved via the CVX software \cite{grant2014cvx}. CVX utilizes the SDP3 solver, which utilizes interior-point methods to solve convex problems. Therefore, the complexity of $\mathcal{P}_{2.1}$, $\mathcal{P}_{2.3}$ and $\mathcal{P}_{3}$ is $\mathcal{O}\big(({2M})^{3.5}\big)$. Thus, the overall complexity of \textbf{Algorithm 2} is $\mathcal{O}\big(I_c({2M})^{3.5}\big)$, where $I_c$ is the number of iterations required for convergence. Likewise, the complexity of \textbf{Algorithm 3} is $\mathcal{O}\big(I_eI_k({2M})^{3.5}\big)$, where $I_e$ and $I_k$ denote number of iterations required for convergence and step 9 satisfaction  of \textbf{Algorithm 3}, respectively. Similarly, the complexity of \textbf{Algorithm 4} is $\mathcal{O}\big[I_d({2M})^{3.5}\big]$, where $I_d$ denotes the number of iterations required for convergence. Meanwhile, the provided numerical results approve that the \textbf{Algorithm 2}, \textbf{Algorithm 3} and \textbf{Algorithm 4} converge in a few iterations.
\section{NUMERICAL RESULTS}
In this section, for the simulation scenario, numerical results are presented to emboss the performance of the proposed system for various examples. we assume that the BS is equipped with a uniform linear array with $N_t=4$ antennas and is deployed at $(0, 0)$. Also, STAR-RIS is placed at $(120m, 0m)$. Moreover, two FD users located in the transmission and reflection region of STAR-RIS, which are capable of sending and receiving signals. The large scale path loss is modeled by $PL=-35.6-10\alpha\log_{10}(d)$ dB, wherein $d$ is relative distance between transmitter-receiver pair and $\alpha$ denotes the path loss exponent. In practice, since STAR-RIS is usually deployed at desirable positions, we assume that the reflected and transmitted signals through STAR-RIS experience less path loss in comparison to the direct signals transmitted from the BS and users. Therefore, the path loss exponents are $\alpha_{B-SR}=2.1$ for the BS-STAR-RIS links, $\alpha_{SR-U}=2.2$ for the STAR-RIS-users links, $\alpha_{B-U}=4$ for the BS-users links and $\alpha_{U-U}=3.1$ for the user-user link. In addition, channels for the STAR-RIS-aided links have been considered Rician fading channels and we have assumed Rayleigh fading channels for the direct links. In particular, the LoS components are modelled by the product of two steering vectors similar to \cite{wu2019intelligent}, \cite{pan2020multicell}. Thus, the small-scale channels for the BS-STAR-RIS links are modelled by Rician fading as follows
\begin{equation}
\boldsymbol{F}= \sqrt{\frac{\kappa}{1+\kappa}}\boldsymbol{b}_M(\vartheta^{\textrm{AoA}})\boldsymbol{b}_{N_t}^H(\vartheta^{\textrm{AoD}})+\sqrt{\frac{1}{1+\kappa}}\boldsymbol{F}^\textrm{NLOS},
\end{equation}
and the small-scale channels for STAR-RIS-users links are modeled as 
\begin{equation}
\boldsymbol{f}= \sqrt{\frac{\kappa}{1+\kappa}}\boldsymbol{b}_M(\vartheta^{\textrm{AoA}})+\sqrt{\frac{1}{1+\kappa}}\boldsymbol{f}^\textrm{NLOS},
\end{equation}
where $\kappa = 4$ represents the Rician factor and $\boldsymbol{b}_M(\vartheta^{AoA})\boldsymbol{b}_{N_t}^H(\vartheta^{AoD})$ denotes the LOS component.
Variables $\vartheta^{AoA}$ and $\vartheta^{AoD}$ represent the angle-of-arrival (AoA) and angle-of-departure (AoD) of STAR-RIS which are uniformly distributed over $\left[0, 2\pi \right)$, respectively.\vfill
The term $\boldsymbol{b}_n \in \mathbb{C}^{n \times 1}$ denotes the steering vector and is defined as 
\begin{equation}
\boldsymbol{b}_n={\Big[1,e^{j\frac{2\pi D }{\lambda_r}\sin{\vartheta}},\dots,e^{j\frac{2\pi D }{\lambda_r}(n - 1)\sin{\vartheta}}\Big]}^T,
\end{equation}
where $D$ is the antenna element separation, $\lambda_r$ is the carrier wavelength and $\frac{D}{\lambda_r}=\frac{1}{2}$ is used.\\
The channels $\boldsymbol{F}^\textrm{NLOS}$ and $\boldsymbol{f}^\textrm{NLOS}$, which  are NLOS component and channels between users are modeled by zero-mean and unit variance Rayleigh distribution RVs. The carrier frequency is assumed to be 3 GHz according to the 3GPP standard used in \cite{bjornson2019intelligent}.\\
However, our simulation is independent of the carrier frequency. Also, the weighting parameters $\alpha_1$ and $\alpha_2$ are assumed to be $0.5$ unless their values are provided. Other simulation parameters are listed in Table \ref{tab1}.
\begin{center}
\begin{table}[H]
\huge
\captionsetup{font=scriptsize}
\captionsetup{justification=centering}
\caption{NUMERICAL RESULTS PARAMETERS}\label{tab1}
\resizebox{\columnwidth}{!}{\begin{tabular}{|l|l|}
\hline
\textbf{Parameters}                      & \textbf{Values} \\ \hline
Maximum transmission power at BS, $P_{max}^{BS}$     & $35 \left[\textrm{dBm}\right]$   \\ \hline
 Maximum transmission power at UL users, $P_{max}^l, \forall l$    & $11 \left[\textrm{dBm}\right]$        \\ \hline
 Noise power at DL users & $-100 \left[\textrm{dBm}\right]$                 \\ \hline
 Noise power at BS  &  $-110 \left[\textrm{dBm}\right]$                \\ \hline
Residual self interference channel variance, ${\hat{\sigma}}^2$  & $-95 \left[\textrm{dBm}\right]$                 \\ \hline
Rician factor for transmitting and reflecting links        & $4 \left[\textrm{dB}\right]$                \\ \hline
 Algorithms convergence parameters,$\big\{\epsilon_1,\epsilon_2,\epsilon_3,\epsilon_4 \big\}$                        &      $\makecell{\big\{10^{-3},10^{-3} ,\\10^{-4},10^{-3}\big\}}$           \\ \hline
\end{tabular}}
\end{table}
\end{center}
\subsection{Convergence}
\begin{figure}[H]
    \centering
    \captionsetup{justification=centering}
    \scalebox{0.6}
    {\includegraphics{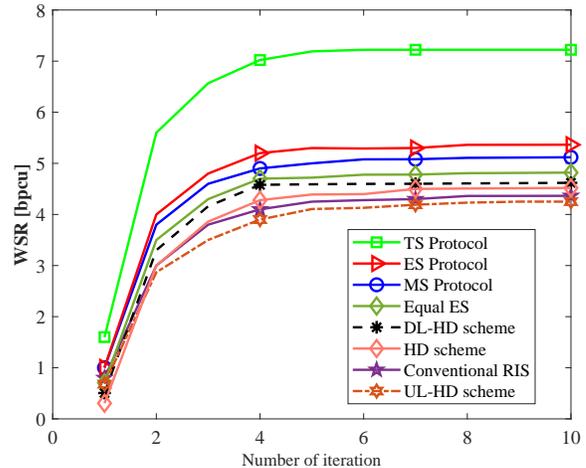}}
     \caption{Convergence of the proposed algorithms.}
    \label{fig:2}
\end{figure}

Convergence of the aforesaid algorithms, i.e, \textbf{Algorithm 2}, \textbf{Algorithm 3} and \textbf{Algorithm 4} are discussed  in Fig. \ref{fig:2}; the WSR versus the number of iterations is depicted for the case of $\big\{N_t = 4, M = 40, K = 2, L = 2, \alpha_1=\alpha_2=0.5\big\}$, where, we compare the proposed approach with several baselines:\\
1) The equal ES method, e.g., setting $\beta_m^t=\beta_m^r=0.5, \forall m \in \mathcal{M}$ and only optimizing the phase; 2) The conventional RIS method, where a transmitting-only RIS and a reflecting-only RIS are adjacent to each other and located at the same coordinate as the STAR-RIS, and each RIS has $M\big /2$ elements \cite{mu2021simultaneously}; 3) The DL-HD scheme, which two FD users have been changed to HD users and capable of receiving signal at DL scenario; 4) The UL-HD scheme, where two FD users have been changed to HD users and capable of transmitting signal at UL scenario; 5) The HD scheme, which a HD user is located at R region and operate at DL mode and also a HD user is located at T region and operate at UL mode.\\
Moreover, the ES protocol was applied for DL-HD scheme, UL-HD scheme and HD scheme. These methods are labelled as \enquote{TS Protocol}, \enquote{ES Protocol}, \enquote{MS Protocol}, \enquote{Equal ES}, \enquote{Conventional RIS}, \enquote{DL-HD scheme}, \enquote{UL-HD scheme} and \enquote{HD scheme}.\\
It is also assumed that the user which deployed at the reflection region of STAR-RIS is randomly and uniformly located in a circle
centred at (120, 5) with a radius of 10 m and the user which deployed at the transmission region is uniformly located in a circle centred at (120, -5) with a radius of 10 m. As it is shown, it can clearly be seen that the proposed algorithms converge rapidly for all methods.
For instance, each method converges in 10 iterations on average.\\
In addition, by increasing the size of STAR-RIS and the BS antennas, it takes more iterations to converge and also leads each iteration to have a higher complexity.    
\subsection{Impact of Number of Elements at STAR-RIS}
\begin{figure}[H]
    \centering
    \scalebox{0.6}
    {\includegraphics{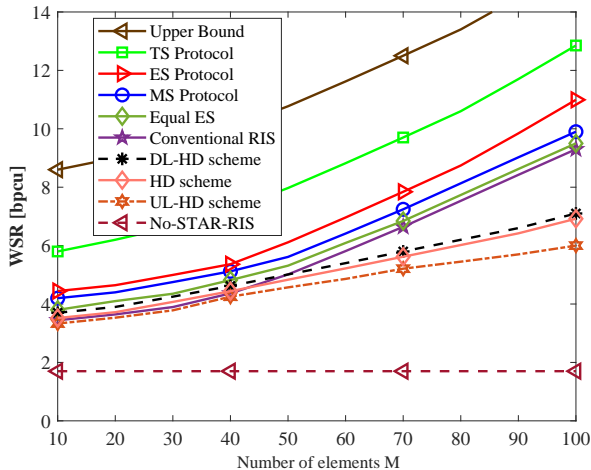}}
    \caption{Weighted sum-rate versus number of STAR-RIS elements for different methods.}
    \label{fig:3}
\end{figure}

Fig. \ref{fig:3} plots the WSR versus the number of STAR-RIS elements $M$. We consider ten scheme; \enquote{TS Protocol}, \enquote{ES Protocol}, \enquote{MS Protocol}, \enquote{Equal ES}, \enquote{Conventional RIS}, \enquote{DL-HD scheme},  \enquote{UL-HD scheme}, \enquote{No-STAR-RIS}, \enquote{HD scheme} and \enquote{Upper Bound} . The HD version of the proposed algorithm namely \enquote{DL-HD scheme} and \enquote{UL-HD scheme} are analyzed for benchmarking, i.e., the DL and UL transmission are performed in two equal time slots. For \enquote{No-STAR-RIS} scheme, no STAR-RIS is adopted, i.e, without utilizing any STAR-RIS and \textbf{Algorithm 1} is also used for optimization.\\
For the sake of comparison, by neglecting the co-channel interference on the DL user caused by the other UL user and also the self-interference at the BS, a performance upper bound is obtained.\\
We can find that by increasing the number of elements of STAR-RIS, the growth in WSR with $M$ for all methods is achieved except the \enquote{No-STAR-RIS} scheme. The three proposed protocols, namely the ES, MS and TS protocol, and even the Equal ES scheme in FD mode, outperform other schemes.\\ Furthermore, for a small number of elements, the DL-HD scheme and HD scheme outperforms convectional RIS but with increasing the number of elements conventional RIS outperforms the two aforesaid schemes.\\
For the sake of HD mode comparison, the DL-HD scheme outperforms the HD scheme and UL-HD scheme by increasing the number of elements. Moreover, it can be seen that by exploiting the BS in FD mode, the gain of increasing $M$ is more beneficial than that of the HD mode.\\
Moreover, for sake of protocol comparison, the TS protocol outperforms the ES protocol and the MS protocol.\vfill
 \subsection{Impact of STAR-IRS Location}
\begin{figure}[H]
    \centering
    \scalebox{0.6}
    {\includegraphics{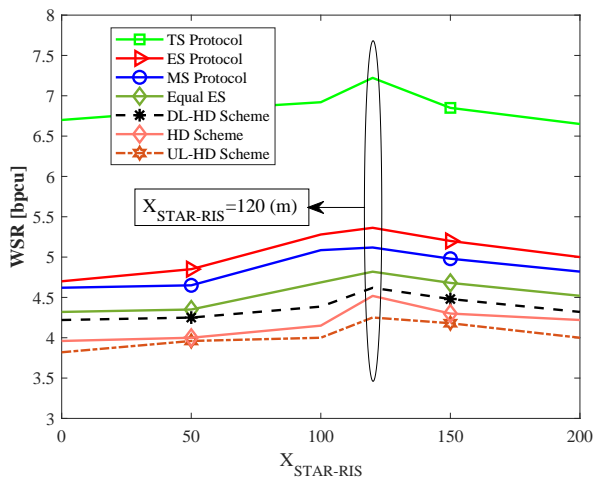}}
    \caption{Weighted sum-rate versus STAR-RIS location.}
    \label{fig:6}
\end{figure}

Fig. \ref{fig:6} investigates the impact of STAR-RIS locations on the WSR for aforesaid protocols and baseline schemes.\\
 It is assumed that the location of $U_1$ is set to $(120, 20)$, which is located in R region, and the location of $U_2$ is set to $(120, -20)$, which is located in T region, for the case of $\big\{N_t = 4, M = 40, K = 2, L = 2, \alpha_1=\alpha_2=0.5\big\}$.\\
 To analyze the impact of STAR-RIS locations in two-dimensional constrained space, it is assumed
that the STAR-IRS is located at $(X_\textrm{STAR-RIS}, 0)$ where $X_\textrm{STAR-RIS} \in \big[0, 200\big]$, i.e., its location can change on the line from (0, 0) to (200, 0). It can be observed that at the location of STAR-RIS, where its location is (120, 0), i.e., $X_\textrm{STAR-RIS}=120(m)$, the WSR of all STAR-RIS protocols and baseline scheme are maximized.\\
Therefore, the STAR-RIS's location for maximizing the WSR is clarified and it's better to be close to users. 
\subsection{Impact of Maximum Transmit Power at DL and UL}
\begin{figure*}
\begin{multicols}{2}
    \includegraphics[width=1\linewidth]{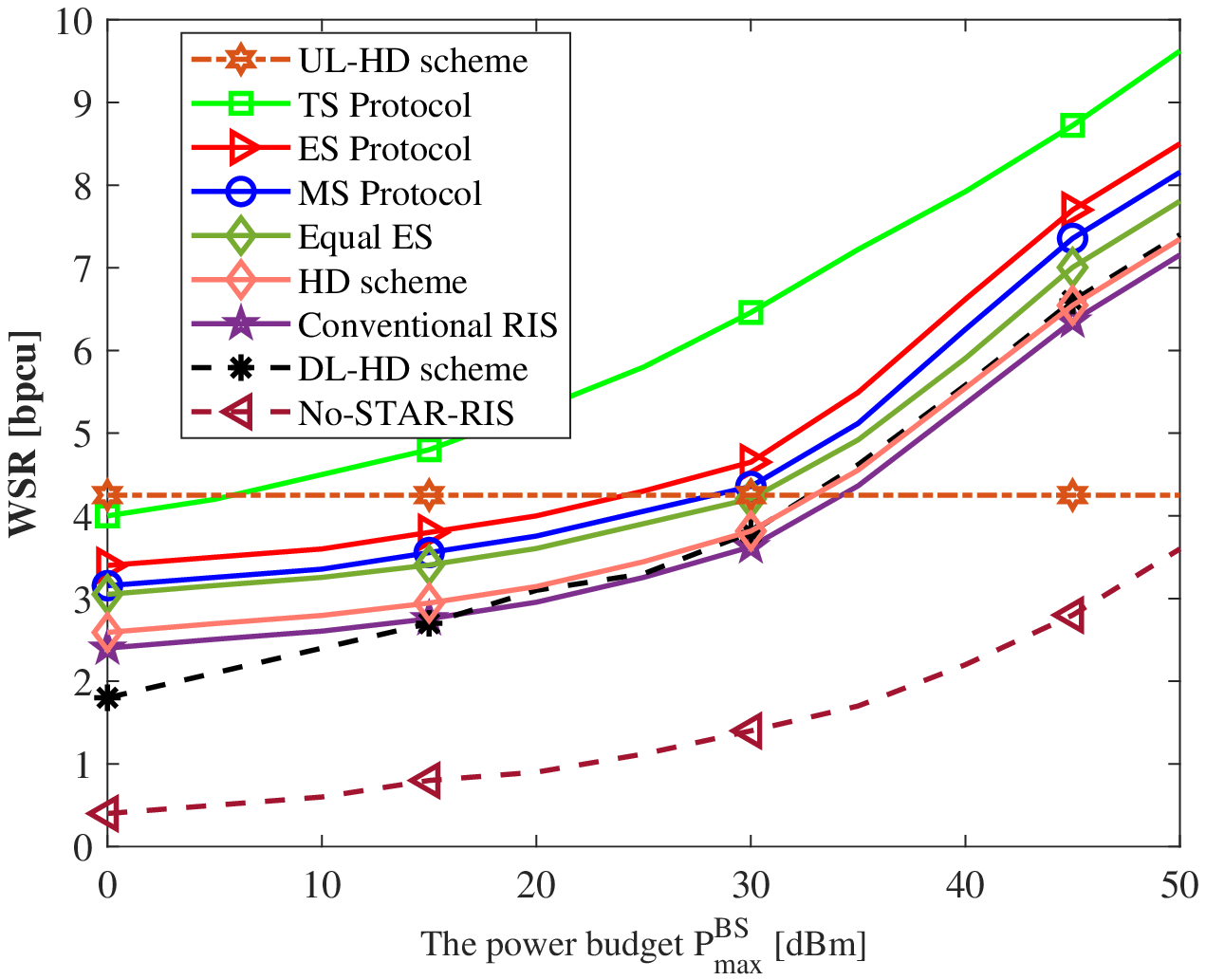} 
    \centering
    
    (a)\par
    
    \includegraphics[width=1\linewidth]{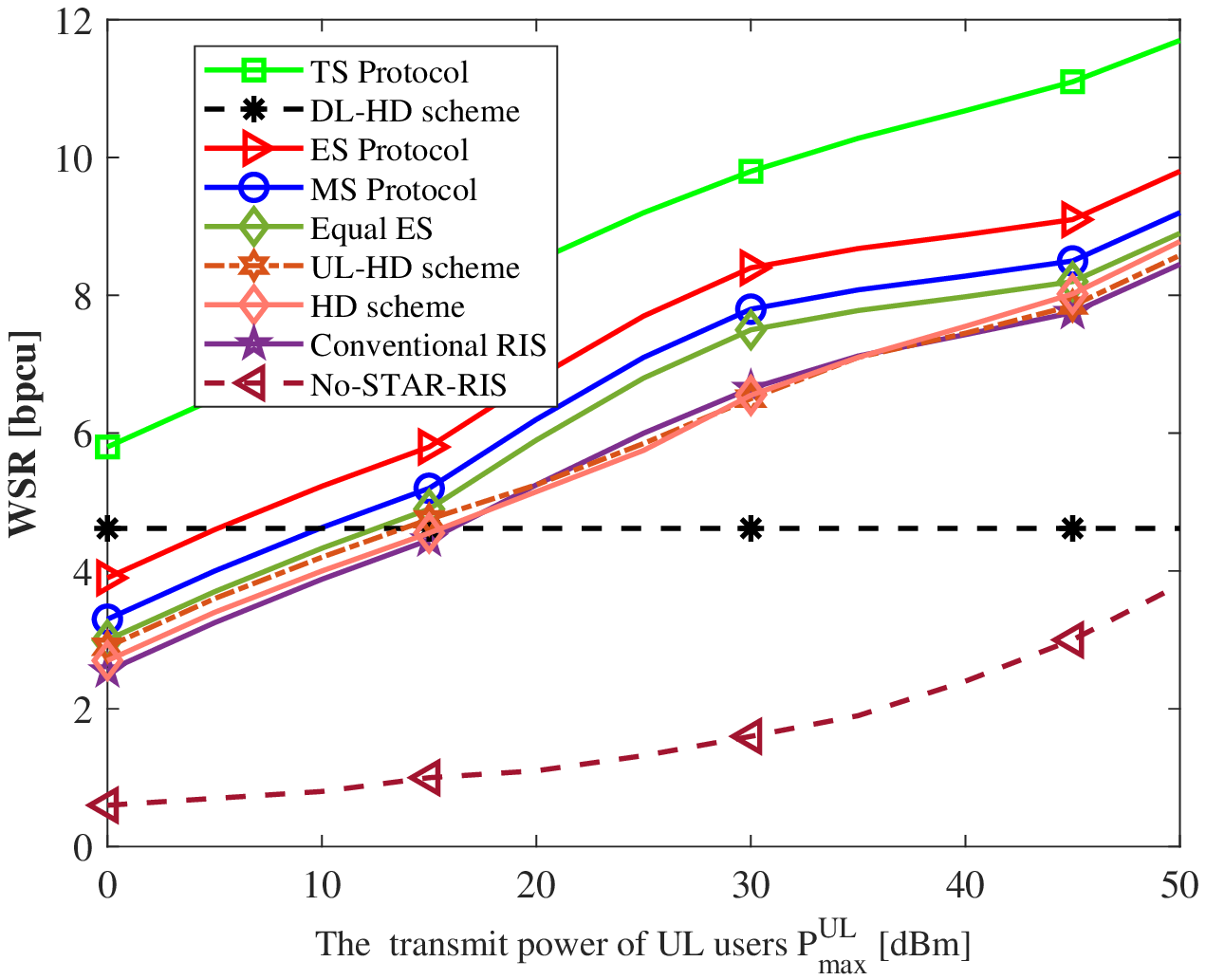}
    \centering
    (a)\par
    \end{multicols}
\begin{multicols}{2}
\includegraphics[width=1\linewidth]{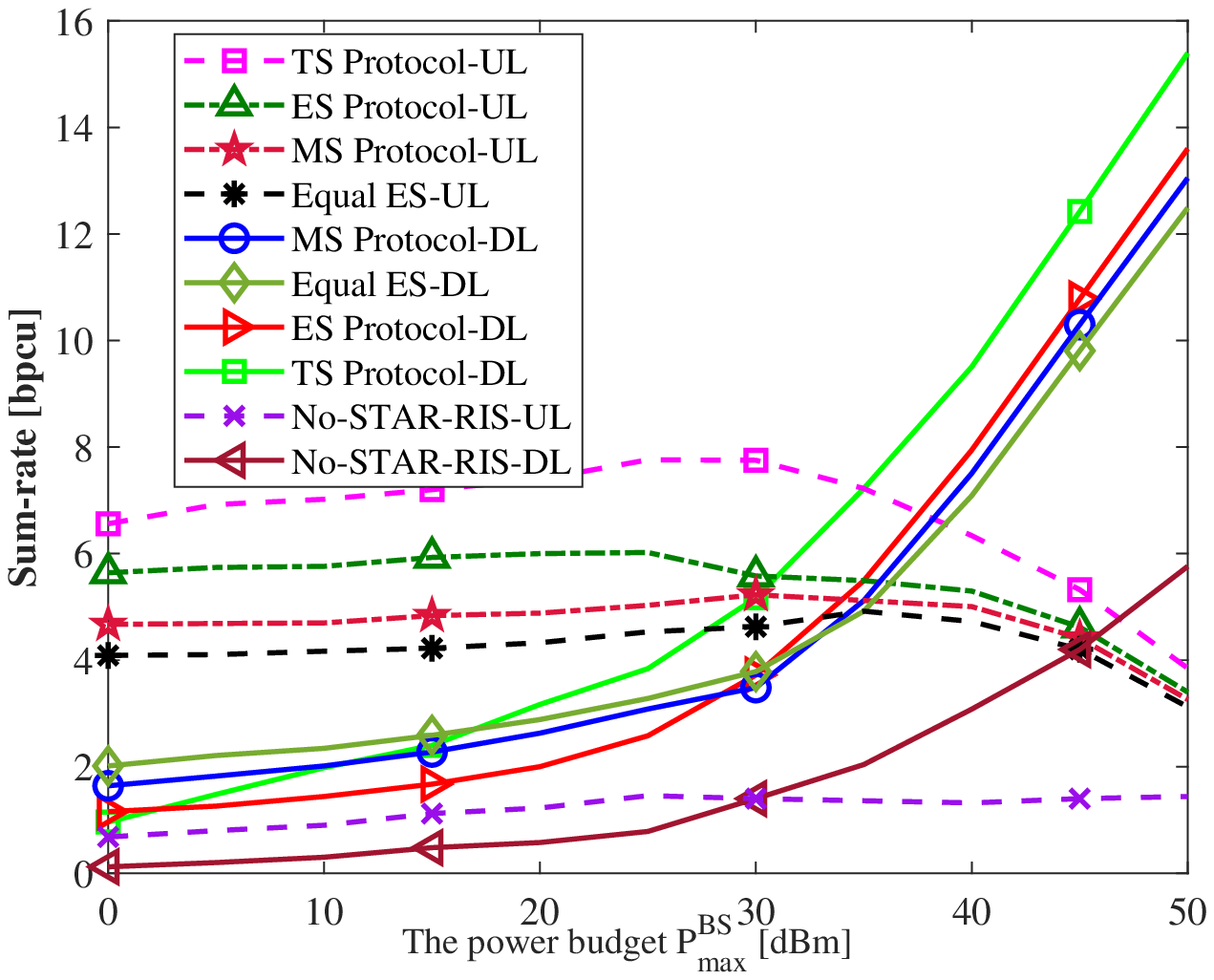}
\centering
(b)\par
\caption{Performance of different schemes for various maximum transmit power at BS, i.e., $P_{max}^{BS}$ for $K=2$, $L=2$, $N_t=4$, $M=40$ and $P_{max}^l= 11 \left[\textrm{dBm}\right]$. (a) WSR versus $P_{max}^{BS}$. (b) Sum-rate versus $P_{max}^{BS}$.}\label{fig:4}\hfill 
\includegraphics[width=1\linewidth]{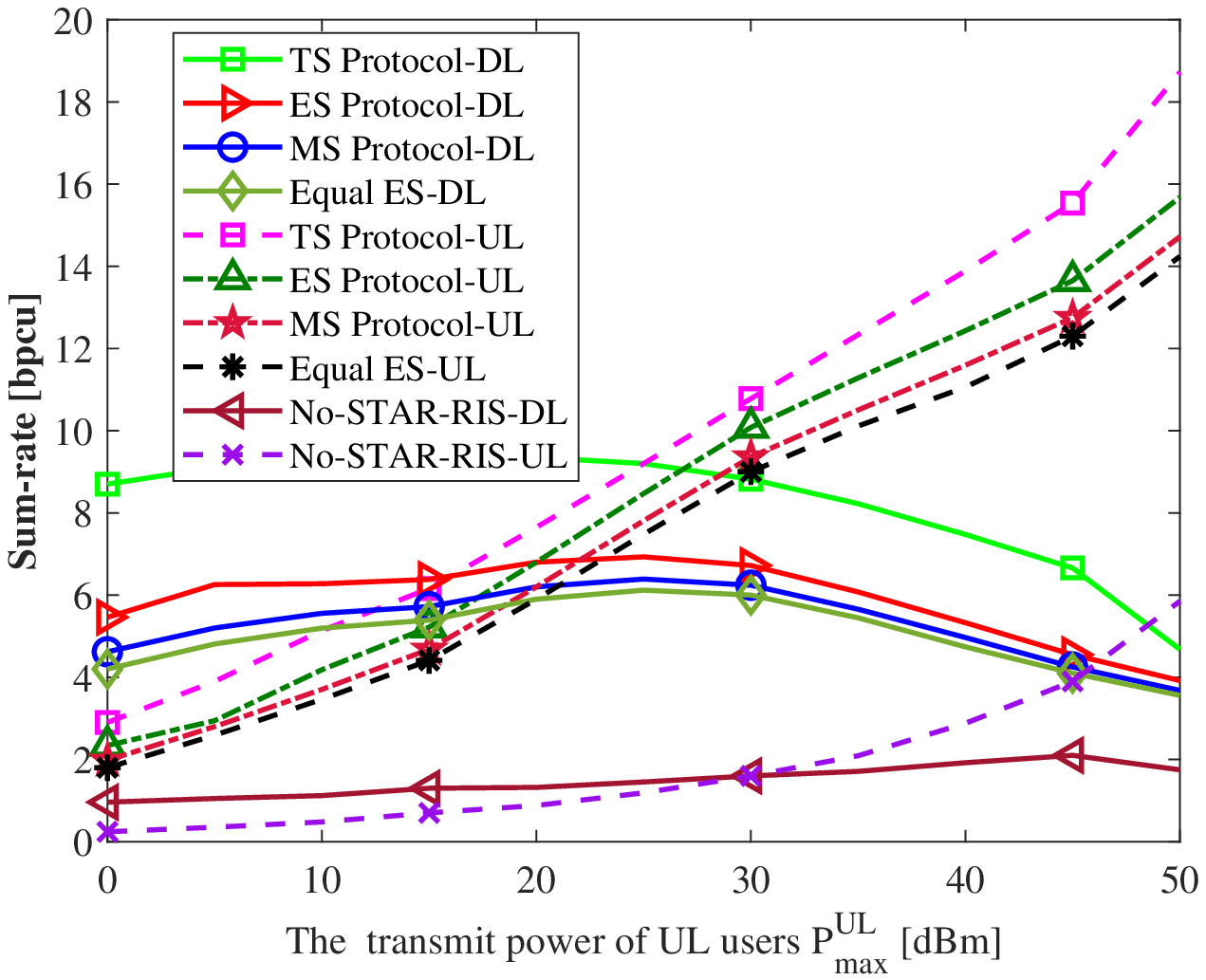}
\centering
(b)\par
\caption{Performance of different schemes for various maximum transmit power at UL users, i.e., $P_{max}^{UL}$ for $K=2$, $L=2$, $N_t=4$, $M=40$ and $P_{max}^{BS}= 35 \left[\textrm{dBm}\right]$. (a) WSR versus $P_{max}^{UL}$. (b) Sum-rate versus $P_{max}^{UL}$.}\label{fig:5}
\end{multicols}
\end{figure*}

Fig. \ref{fig:4}a illustrates the WSR as a function of the the maximum transmit power of the BS, i.e., $P_{max}^{BS}$ and performance of three  protocols of the STAR-RIS with FD scenario and HD schemes are compared for $K = 2$, $L = 2$, $N_t = 4$, $M = 40$ and $P_{max}^l= 11 \left[\textrm{dBm}\right]$.\\
We can see that the WSR increases with $P_{max}^{BS}$ for all schemes except the UL-HD scheme, while the TS protocol achieves higher WSR than other methods, especially in the high values of the maximum power of the BS, since it can eliminate the inter-user interference.\\
In the UL-HD scheme, the two users operate at UL mode and the self-interference power at the BS is ignored and the WSR in the aforesaid scheme remains constant and disproportionate   
 to $P_{max}^{BS}$ and outperforms all schemes in low values of the maximum power of the BS.\\
 Furthermore, STAR-RIS achieves better performance than conventional RISs, since, with a given number of elements for utilizing conventional RISs for the two modes, the  the degrees of freedom can not be fully exploited.\\
 Even, the HD scheme and DL-HD scheme outperform conventional RIS and in the high value of the maximum power of the BS, conventional RIS outperforms the UL-HD scheme.\\
 Fig. \ref{fig:4}b  also indicates the effect of $P_{max}^{BS}$ on the sum rate of UL and DL data transmissions for protocols of STAR-RIS. Although the self-interference power degrades the uplink performance, it is shown that the WSR improves monotonically by increasing the maximum transmit power of the BS.\\
 Eventually, to maximize the WSR for high values of the maximum power of the BS, it can clearly be seen that the UL users do not participate in the system performance.\\
 In contrast to the sum-rate of downlink, as anticipated, the sum-rate of the uplink degraded by increasing the maximum transmit power of the BS due to increasing the power of the self-interference.\\
 Fig. \ref{fig:5} investigates the effect of the maximum transmit power of UL users on the WSR and sum rates of UL and DL. Similarly, for aforesaid schemes with FD and HD scenarios, the achievable rates versus $P_{max}^{UL}$ is illustrated for the case of $K = 2$, $L = 2$, $N_t = 4$, $M=40$ and $P_{max}^{BS}=35 \left[\textrm{dBm}\right]$. According to Fig. \ref{fig:5}a,  the WSR increases with $P_{max}^{UL}$ for all schemes except the DL-HD scheme, while the TS protocol achieves higher WSR than other methods.\\
 In the DL-HD scheme, the two users operate at DL mode and uplink user-interference is not relevant to this scheme, therefore, the WSR in the aforesaid scheme remains constant and disproportionate to $P_{max}^{UL}$ and outperforms other methods in low value of the maximum transmit power of UL users, but, by increasing the $P_{max}^{UL}$ other methods outperform DL-HD scheme.\\
 In addition, STAR-RIS achieves better performance than conventional RIS and even HD scheme and UL-HD scheme outperforms conventional RIS in the high value of $P_{max}^{UL}$. Furthermore, STAR-RIS protocols outperform other baseline schemes.\\ Moreover, the downlink and uplink sum rates for STAR-RIS protocols are indicated in Fig. \ref{fig:5}b; the downlink sum-rate degrades by increasing $P_{max}^{UL}$, and in contrast,  the uplink sum-rate increases in $P_{max}^{UL}$.
%%====> 5. Conclusion <===%%
\section{Conclusion} \label{Sec:Sec5}
In this paper, we investigated the effect of deploying STAR-RIS and utilizing various protocols in a FD communication system. Particularly, the concentration was on joint optimization for STAR-RIS amplitude and phase shift, the beamformer and combining vectors at the BS, and the transmitted power of the uplink users. The WSR maximization problem subject to the maximum power constraints at the BS and the uplink users was considered, and an iterative approach was proposed. We coped with the problem by utilizing the alternative optimization method wherein the WMMSE approach was also used due to the non-concavity of optimization problem. We firstly transformed the optimization problem into several convex sub-problems and solve them separately in an iterative procedure to achieve optimal solutions. Moreover, in our proposed algorithm, the optimized amplitude and phase shifts of STAR-RIS were obtained via the SCA technique. The complexity of the overall proposed algorithm was discussed, and its convergence was verified through numerical results. Finally, the effects of the transmission power of the BS and the uplink users, the size and the location of STAR-RIS were discussed and compared for various baseline schemes to clarify performance increment of the proposed algorithm. In addition, by using optimized STAR-RIS, the WSR is improved when the users are FD users and capable of transmitting and receiving signals, simultaneously. It is concluded that using optimized STAR-RIS in a FD scenario is more beneficial than using a conventional RIS.     
%%====> References <===%%
\bibliographystyle{IEEEtran}
\bibliography{STAR-RIS} 
\end{document}